%% file: paper.tex
\documentclass[10pt,sigconf,letterpaper,nonacm]{acmart}

\usepackage[english]{babel}
\usepackage{blindtext}

\renewcommand\footnotetextcopyrightpermission[1]{} 
\setcopyright{none}

\acmDOI{}

\acmISBN{}

\acmConference[]{}
\acmYear{2022}
\copyrightyear{}

\acmPrice{}

\usepackage{microtype}
\usepackage{graphicx}

\usepackage[position=bottom]{subfig}
\usepackage[font=small,format=plain,labelfont=bf,textfont=it]{caption}
\usepackage[title]{appendix}
\usepackage{booktabs}
\usepackage{tabularx}
\usepackage{tabu}
\usepackage{siunitx}
\usepackage{multirow}
\usepackage{tcolorbox}
\usepackage{xcolor}
\usepackage{color, colortbl}
\usepackage{balance}
\usepackage{placeins}
\usepackage{url}
\usepackage{amsmath}
\usepackage{bbding}
\usepackage[htt]{hyphenat}
\usepackage[compact]{titlesec}
\usepackage{cleveref}

\crefformat{section}{\S#2\color{blue}#1#3} 
\crefformat{subsection}{\S#2\color{blue}#1#3}
\crefformat{subsubsection}{\S#2#1#3}

\titlespacing\section{0pt}{6pt plus 4pt minus 2pt}{0pt plus 2pt minus 2pt}
\titlespacing\subsection{0pt}{6pt plus 4pt minus 2pt}{0pt plus 2pt minus 2pt}
\titlespacing\subsubsection{0pt}{6pt plus 4pt minus 2pt}{0pt plus 2pt minus 2pt}

\renewcommand{\paragraph}[1]{\vspace*{0.03in}\noindent\textbf{#1}}

\newcommand{\eg}{{\it e.g.}}
\newcommand{\ie}{{\it i.e.}}

\newcommand{\code}[1]{\texttt{#1}}
\newcommand{\dataset}{{\textsf Fixed Dataset}\xspace}
\newcommand{\datasetlarge}{{\textsf Evolving Dataset}\xspace}

\newcommand{\takeaway}[1]{\textcolor{cyan}{}}

\makeatletter
\def\input@path{{./sections/}}
\makeatother

\begin{document}
\title[LEAF]{LEAF: Navigating Concept Drift in Cellular Networks}

\author{
{\rm Shinan Liu, Francesco Bronzino$^\dagger$, Paul Schmitt$^\ddagger$, Arjun Nitin Bhagoji, \\ Nick Feamster, Hector Garcia Crespo$^\mathsection$}, {Timothy Coyle\rm $^\mathsection$, Brian Ward$^\mathsection$}\\
\textit{University of Chicago, $^\dagger$École Normale Supérieure de Lyon, \\ 
$^\ddagger$University of Hawaii, $^\mathsection$Verizon} \\
\normalsize
\textit{\{shinanliu, abhagoji, feamster\}@uchicago.edu, francesco.bronzino@ens-lyon.fr, pschmitt@hawaii.edu,} \\
\textit{\{hector.garcia, timothy.coyle, brian.ward2\}@verizon.com}\\
}

\renewcommand{\shortauthors}{}

\begin{sloppypar}

\input{abstract}

\maketitle

\input{introduction}
\input{context}

\input{characterization}

\input{leaf}

\input{casestudy}
\input{evaluation}
\input{related}
\input{conclusion}

\newpage
\bibliographystyle{ACM-Reference-Format}
\bibliography{paper}

\end{sloppypar}

\setcounter{page}{1}
\input{appendix}

\end{document}

%% file: sections/abstract.tex
\begin{abstract} 
    Operational networks commonly rely on machine learning models for many
    tasks, including detecting anomalies, inferring application performance, and
    forecasting demand. Yet, model accuracy can degrade due to {\em concept
    drift}, whereby the relationship between the features and the target to be
    predicted changes. Mitigating concept drift is an essential part of
    operationalizing machine learning models in general, but is of particular
    importance in networking's highly dynamic deployment environments. In this
    paper, we first characterize concept drift in a large cellular network for a
    major metropolitan area in the United States. We find that concept drift
    occurs across many important key performance indicators (KPIs),
    independently of the model, training set size, and time interval---thus
    necessitating practical approaches to detect, explain, and mitigate it. We
    then show that frequent model retraining with newly available data is not
    sufficient to mitigate concept drift, and can even degrade model accuracy
    further. Finally, we develop a new methodology for concept drift mitigation,
    Local Error Approximation of Features (LEAF). LEAF works by detecting drift;
    explaining the features and time intervals that contribute the most to
    drift; and mitigates it using forgetting and over-sampling. We evaluate LEAF
    against industry-standard mitigation approaches (notably, periodic
    retraining) with more than four years of cellular KPI data. Our initial
    tests with a major cellular provider in the US show that LEAF consistently
    outperforms periodic and triggered retraining on complex, real-world data
    while reducing costly retraining operations.
\end{abstract}


%% file: sections/introduction.tex
\section{Introduction}\label{sec:intro}

Network operators rely on machine learning (ML) models to perform many tasks,
including anomaly detection~\cite{shon2007hybrid, shon2005machine}, performance
inference~\cite{futuriom2021Verizon} and diagnosis, and
forecasting~\cite{mei2020realtime, chinchali2018cellular}. However, deploying and
maintaining these models can prove challenging in
practice~\cite{sommer2010outside}. One significant operational challenge is {\em
concept drift}, wherein a model that is initially accurate becomes less accurate
over time due to sudden changes, periodic variation, or gradual drift.
Previous work in applying ML models to network management tasks
has generally trained and evaluated models on fixed, offline
datasets~\cite{sinclair1999application, dong2016comparison, shon2007hybrid,
shon2005machine, ayoubi2018machine, mei2020realtime, chinchali2018cellular},
demonstrating the ability to predict various network features at
fixed points in time on a static dataset. Yet, a model that performs well
offline on a single dataset may not in fact perform well in practice, especially
over time as characteristics change.


Concept drift is a relatively well-understood phenomenon in ML for other 
prediction problems (\eg, image and text classification~\cite{schlimmer1986incremental,
lu2018learning,gama2014survey}). 
But mitigating concept drift for networking problems, particularly forecasting 
Key Performance Indicators (KPIs) in a cellular network introduces fundamentally new challenges that make previous
approaches from other domains inapplicable. Networks have unique
characteristics, such as dynamic signal interference due to environment
changes (\eg, changes due to weather, seasonality,
etc.)~\cite{zheng2014rethinking}, which calls for new approaches.  
In contrast to previous tasks, 
where the semantics of prediction occurs on a fixed object and characteristics 
of the features change relatively slowly over time~\cite{fdez2007applying, vzliobaite2010adaptive, gama2014survey},
predictions of network characteristics occur continuously, and occur within 
the context of a system that changes over time 
due to periodicity (\eg, seven-day period of volume), gradual evolution (\eg, the constant addition of capacity by new equipment 
installations), and exogenous shocks (\eg, a software upgrade, or a sudden change in traffic
patterns or demands such as the COVID-19 pandemic, which resulted in significant 
changes due to factors like reduced mobility~\cite{lutu2020characterization} or 
user behaviors~\cite{liu2021characterizing}). 

Beyond simply detecting concept drift, operators often want to
explain why a model has become less accurate and mitigate it. Thus, the ability to 
diagnose the behavior of black-box regression models is necessary to help
operators use these models in practice. To this end, this paper develops
(1)~approaches to explain how different features contribute to
drift; and (2)~a strategy to mitigate drift through triggered, focused
re-retraining, forgetting, and over-sampling approaches.  
Although past work has developed methods to help explain concept drift, it has largely 
focused on classification problems~\cite{jordaney2017transcend,yang2021cade}; 
in contrast, prediction problems in the context of cellular networks are 
often regression problems.

Overall, this paper makes three contributions:

First, {\bf we characterize concept drift in the context of a large cellular
network}, exploring drift for KPIs using more 
than four years of KPI data in a major United States city
and surrounding metropolitan area, from one of the largest
cellular providers in the United States.
We demonstrate concept drift in a large cellular network
comparing different KPIs, model families, training set sizes, and periods
across many regression models and tasks. 
Concept drift occurs consistently and independently of both the 
size and period of the training set. The diverse, longitudinal 
nature of the dataset used presents a challenging concept drift problem.

To detect, mitigate and explain this concept drift in cellular networks, {\bf we introduce {\em LEAF} (Local Error Approximation of Features)} as an explainable ML approach for networking models. We apply
Kolmogorov-Smirnov Windowing (KSWIN) to time-series of estimated errors, which
tells us when a model drifts. We find the most representative
features that reflect error distributions, and use LEAplot and LEAgram
to inform operators about for \textit{which features}, \textit{where}, and 
\textit{how much} concept drift occurs, which maps
to the under-trained region of a model. Based on these explanations, the
framework strategically re-samples the training data, and creates
temporal ensembles to mitigate drift. In addition to providing explanations 
for drift, LEAF also outperforms the baseline approach of regular 
retraining using a fixed-size window of recent observations, which 
we find can even reduce model performance in a number of settings!

Finally, {\bf we evaluate the effectiveness and efficiency of LEAF holistically on the complete dataset and showcase the 
explanation power}, 
across boosting, bagging, and LSTM-based models and a variety of forecasting targets.
LEAF consistently outperforms existing mitigation approaches, while reducing costly
retraining operations by as much as 76.9\% (compared to periodic retraining). 
While KPIs with higher dispersion are harder to mitigate, 
LEAF is still able to consistently reduce errors. From a case study, 
LEAF provides possible explanations of drift by identifying the most responsible 
features. LEAplot further localizes the high-error eNodeBs, where the 
top 5\% of error mostly comes from suburban areas. From LEAgram, we also find 
that the over and underestimation changes over time, and the overestimation 
phase correlates highly with less mobility.

To the best of our knowledge, LEAF is the first method to provide explanations 
for concept drift for regression, building on previous work that has largely 
focused on classification problems~\cite{jordaney2017transcend,yang2021cade}.
We thus envision LEAF being applied beyond cellular 
networks, to other network management problems that use black-box models for 
regression-based prediction. Our belief in the 
generalizability of LEAF arises from the diversity in time series across KPIs 
and time that LEAF tackles successfully.  Such avenues 
present rich opportunities for future work.


%% file: sections/context.tex
\section{Problem Setup}\label{sec:data} 
The main problem we tackle in this paper is that of accurate longitudinal
prediction in the presence of concept drift. This is a challenging problem that
requires datasets with sufficient length and variation to determine if our
proposed methods would be effective in a real-world setting. In this section, we
describe the diverse, longitudinal dataset used in this paper. We then
define our time-series regression problem, which is to predict the current value
of these variables given historical data. Finally, we detail the metrics we use
throughout the paper to measure performance.



\subsection{Dataset Description}\label{subsec:dataset}

Our analysis in this paper is based on more than four years (January 1,
2018 to March 28th, 2022) of daily measurements of LTE cellular network performance 
indicators collected at the eNodeB-level (evolved NodeBs, or the ``base 
station'' in the LTE architecture) from a major wireless carrier in the United States. 
Table~\ref{tab:data} summarizes the characteristics of the datasets, which we refer 
to as \dataset (the dataset that contains a fixed number of eNodeBs each day) and 
\datasetlarge (an expanded set of \dataset with a growing number of eNodeBs) 
in the remainder of the paper. 

\begin{table}[t]
  \centering
  \small
  \begin{tabu}to 0.47\textwidth{r X}
  \toprule
  \textbf{Collection period} & Jan. 1st 2018 --  March 28th 2022 \\ \midrule
  \textbf{Identifiers} & eNodeB ID \& Time stamp \\ \midrule
  \textbf{Number of KPIs} & 224 \\\midrule
  \textbf{Groups of KPIs} & Resource utilization 
  \newline Network performance
  \newline User experience \\ \midrule
  \textbf{Number of eNBs} & \dataset: 412 common eNBs 
  \newline \datasetlarge: 898 eNBs \\ \midrule
  \textbf{Number of logs} & \dataset: 699,381
  \newline \datasetlarge: 1,084,837  \\ \bottomrule
  \end{tabu}
  \caption{Summary of datasets.}
  \label{tab:data}
\end{table}

\noindent \textbf{Why these datasets?} The first requirement for any dataset 
used to analyze concept drift is that it should actually contain clear evidence 
of such drift. Our analysis in Section~\ref{sec:drift} establishes that the variables, 
which are Key Performance Indicators (KPIs) from the cellular network, in \dataset 
and \datasetlarge do indeed drift over time. Intuitively, this drift occurs due 
to endogenous reasons like changes in network infrastructure and KPI definitions, 
as well as exogenous ones such as the COVID-19 pandemic.

Further, \dataset is affected by factors from software upgrades to mobility pattern 
changes. It provides "apples to apples" comparison across 
time. It is diverse since different KPIs exhibit drastically 
different behavior over time. This diversity and the real-world nature of the dataset lead to generalizable insights 
from our methods, since a wide variety of possible time-series behavior is captured 
within our dataset. \datasetlarge extends the KPI diversity further by including the 
operational growth of eNodeBs in this area, adding more heterogeneity from practice. 

\noindent \textbf{Further details.} \dataset contains information from 412
common eNodeBs across time. It is collected in a large city and surrounding 
metropolitan area (rural, suburban, and urban included) in the United States. 
The dataset spans more than four years---from January 1, 2018 to March 28, 
2022---and contains 699,381 daily eNodeB-level logs. \datasetlarge has the 
same information, but with a maximum of 898 eNodeBs, containing 1,084,837 
daily logs. 

Each log contains 224 Key Performance Indicators (KPIs) collected for a base
station on a particular date.  KPIs are statistics collected and used by the operator of the network to monitor and assess network
performance. The 224 KPIs fall into three categories:
(1)~resource utilization (\eg, data volume, peak active users, active 
session time, cell availability rate), (2)~access network performance 
(\eg, throughput, connection establishment success, congestion, packet loss), 
and (3)~user experience features (\eg, call drop rate, RTP gap duration ratio, 
abnormal UE releases).  Further, some of the KPIs 
have separate directional measurements.


  
\subsection{Modeling Goal}\label{subsec:problem}

\noindent \textbf{Forecasting Problem.} We focus our study of concept drift in the context of network 
forecasting. Network forecasting (capacity, performance, user experience) is an 
important problem for operators as it guides infrastructure configuration, management, 
and augmentation.
We focus on per-eNodeB level KPI forecasting to provide suggestions for
capacity adjustment, deployment, maintenance, and operation in large cellular
networks. 

We use historical data---\ie,
all available KPIs and dates (as features) up to a given day---to forecast one
or more target KPIs of interest 180 days in the future. We use a 180-day
forecast window because the model outputs we focus on are used {\em in
practice} for network infrastructure provisioning and augmentation over long
timescales. For such tasks, operators need at least 180 days to plan, decide, 
prepare, and execute capacity augmentation. This 180-day gap also makes it more 
challenging to explain and mitigate drift.

The nature of this problem is regression with time-series information. Regression 
models are a better fit than classification because (1)~we aim to forecast target 
KPIs that have wide ranges of possible numerical values and 
(2)~fine-grained forecasting numerical better enables operators to understand the KPIs and 
take action on their network.

\begin{table}[t!]
  \centering
  \small
  \begin{tabularx}{0.43\textwidth}{ccccccc}
    \toprule
    \multirow{3}{*}{\textbf{Property}} & \multicolumn{2}{c}{\textbf{Resource}}  
    & \multicolumn{2}{c}{\textbf{Network}} & \multicolumn{2}{c}{\textbf{User}}  \\ 
    & \multicolumn{2}{c}{\textbf{Utilization}}  
    & \multicolumn{2}{c}{\textbf{Performance}} & \multicolumn{2}{c}{\textbf{Experience}}\\
    \cmidrule(ll){2-3} \cmidrule(ll){4-5} \cmidrule(ll){6-7}
    & DVol & PU & DTP & REst   & CDR  & GDR  \\ \midrule
    Std/Mean & 0.81 & 1.76 & 0.59 & 0.85 & 2.48  & 8.52 \\ 
    \rowcolor{lightgray}
    Periodic & \Checkmark & \Checkmark & \Checkmark & \Checkmark & \Checkmark & \Checkmark \\ 
    Bursty &  & \Checkmark &  &  & \Checkmark & \Checkmark \\ 
    \rowcolor{lightgray}
    Data Lost &  & \Checkmark &  &  &  & \\ 
    Balanced & & \Checkmark &  & \Checkmark &  & \\ \bottomrule
  \end{tabularx}
  \caption{Characteristics of target KPIs in \datasetlarge. DVol: Downlink volume; 
  PU: Peak active UEs; DTP: Downlink Throughput; REst: RRC establishment success; 
  CDR: S1-U call drop rate; GDR: RTP gap duration ratio. Those of \dataset can 
  be found in Appendix~\ref{subsec:appedix_data}.}
  \label{tab:target_KPIs}
\end{table}

\noindent \textbf{Forecasting Targets.} 
We choose the forecasting targets based on their measurement goals.
We focus on the prediction of six KPIs out of the available 224, two out of 
each of the three KPI groups that are 
most relevant to network planning: measurements of resource utilization
(downlink data volume, peak number of active UEs / User Equipment),
network performance (downlink throughput, RRC / Radio Resource Control 
establishment success),  and user experience (S1-U / S1 User plane external 
interface call drop rate, RTP gap duration ratio). They not only cover 
the RAN events and wireless connections, but also the UE behaviors.

Additionally, the different focus of KPIs also makes them have different statistical 
behaviors. Table~\ref{tab:target_KPIs} summarizes the main characteristics of 
these KPIs in both \dataset and \datasetlarge.
These KPIs exhibit a variety of statistical patterns and characteristics which,
as we will see in Section~\ref{sec:drift} and \ref{sec:evaluation} can ultimately affect how models
drift over time, as well as the best strategies for mitigating drift for a
particular KPI. For \dataset, all KPIs exhibit 7-day periodicity, although some KPIs
exhibit far more variance than others. {\em Downlink volume (DVol)} and {\em RRC establishment
success (REst)} present similar ranges in their distributions.  In contrast,
{\em S1-U call drop rate (CDR)} and {\em peak active UEs (PU)} show a more
bursty behavior over time.  While the data distributions of {\em downlink
throughput (DTP)} and PU have balanced distributions that do not present a long
tail, DVol, REst, and CDR present more skewed distributions. Some data for PU
was lost between July 2019 and January 2020. For comparison, \datasetlarge has 
noticeably higher dispersions (Std/Mean) on PU, CDR, and GDR, due to increased 
infrastructure in this dataset. Moreover, the distributions of DVol and DTP in it 
are more skewed than in \dataset.

\subsection{Metrics for Evaluating Drift}\label{subsec:metrics}

The characterization of model performance over time is essential to evaluate
concept drift. Given that we model network capacity forecasting as a regression
problem, we test model effectiveness by computing distances between prediction
and ground truth by date, specifically using Root Mean Squared Error (RMSE). To
better understand drift across different KPIs whose natural operating value
ranges are drastically different (\eg, call drop rates are scalars mostly less
than 1, while downlink volume scalars are often greater than 300,000), we
normalize the RMSE by maxmin and derive the Normalized Root Mean Squared Error (NRMSE).

In addition to offering an equitable comparison across multiple KPIs, the NRMSE
is well-suited to understand concept drift because (1)~it penalizes large
errors that can be costly for ISPs' operations (\eg, over-provision of
resources); and (2)~it captures the relative impact of errors over extended
periods of time. In practice, NRMSE scores under 0.1 and $R^{2}$ over 90\%
indicate that the regression model has very good prediction
power~\cite{nau2014notes}. \footnote{To avoid possible errors in interpretation, we
also test the performance of regression by the other metrics including
coefficient of determination (\ie, $R^{2}$), mean absolute percentage error,
mean absolute error, explained variance score, Pearson correlation, mean
squared error, and median absolute error. We omit these metrics for brevity. We
observe that all phenomena we describe in the paper using NRMSE
also hold for these metrics.}

\paragraph{Average NRMSE distance from a static model.} To show the long-term effectiveness of different
mitigation schemes, we study the relative evolution of errors over extended
periods of time. For each day in the dataset, we compute the average NRMSE
across all eNodeBs and compare it to the error generated by a model that is
never retrained, \ie, a static model. We define $\Delta \overline{NRMSE}$ as the
average distance between the error for a mitigated model $M_1$ against the
static model $M_0$ in the form of a percentage distance. We define:
\begin{equation}\label{eqt:delta_NRMSE}
    \Delta \overline{NRMSE}(M_1, M_0) = \frac{\overline{NRMSE}(M_1) - 
    \overline{NRMSE}(M_0)}{\overline{NRMSE}(M_0)} \times 100\%
\end{equation}
where $\overline{NRMSE}$ is the average over time.

%% file: sections/characterization.tex
\section{Drift Characterization}\label{sec:drift} 

In this section, we explore how trained forecasting models are affected by concept drift. We
train a number of models from different categories of regression-based
predictors, targeting different KPIs, and different training parameters. 
After identifying concept drift behavior in the cellular network dataset 
we use, we evaluate the effectiveness of periodic retraining with 
recent data, which is the state of the art for drift mitigation. We 
demonstrate that it is hard to identify a single retraining strategy 
across models and KPIs, making na\"ive retraining a difficult strategy 
to implement in practice. All measurements in this section use the \datasetlarge.

\subsection{Model Selection} 
To explore the performance of different widely-adopted regression techniques, we
use for all experiments in this paper the AutoGluon~\cite{autogluon2021} 
pipeline and TensorFlow. AutoGluon is used to quickly prototype deep learning
and machine learning algorithms on various existing frameworks. We 
emphasize that the goal of our study is not to create and tune models 
that maximize performance, but rather to demonstrate and better understand 
concept drift in a real-world network.

We select four different families of models: (1)~gradient boosting algorithms
like LightGBM, LightGBMLarge, LightGBMXT, CatBoost, and XGBoost; (2)~bagging
algorithms such as Random Forest and Extra Trees; (3)~distance-based algorithms
like KNeighbors; (4)~recurrent neural networks like LSTM. All models either 
incorporate temporal features (\eg, time stamps, day of the week, month, year), 
or are time-series models (LSTM). Although it is viable to fine tune each 
model's hyperparameters by hand, we rely on the auto-selection pipeline 
with the goal of a fair comparison and to make training scalable 
and efficient.

For all experiments, we develop models for each selected target KPI. 
As the input of the models, we use a portion of the history of all categorical 
and numerical KPIs from all eNodeBs up to the date 
when the model is generated. we create a single model capable 
of forecasting values for each individual base station. 

\subsection{Comparing Drift Behavior across KPIs}\label{subsec:different}

We explore whether we can observe drift for the KPI forecasting
task and additionally whether the drift patterns from different KPIs are
similar. We train all the models mentioned above for each target KPI, while 
keeping the inputs of models unchanged. We use a 90-day window of historical data from
all eNodeBs with an end date of July 1, 2018 for our training data, \ie, forecasting
KPIs starting from December 28th, 2018 (we plot from Mid-March 2019 because of data 
losses between January and March 2019). We then test these models on data subsets 
split by date. Note that the number of samples evaluated on each date might be different,
because of natural expansions of infrastructure in cellular networks. 
\begin{figure}[!t]
  \centering
  \subfloat[Volume.\label{fig:volume}]{%
    \includegraphics[width=0.245\textwidth]{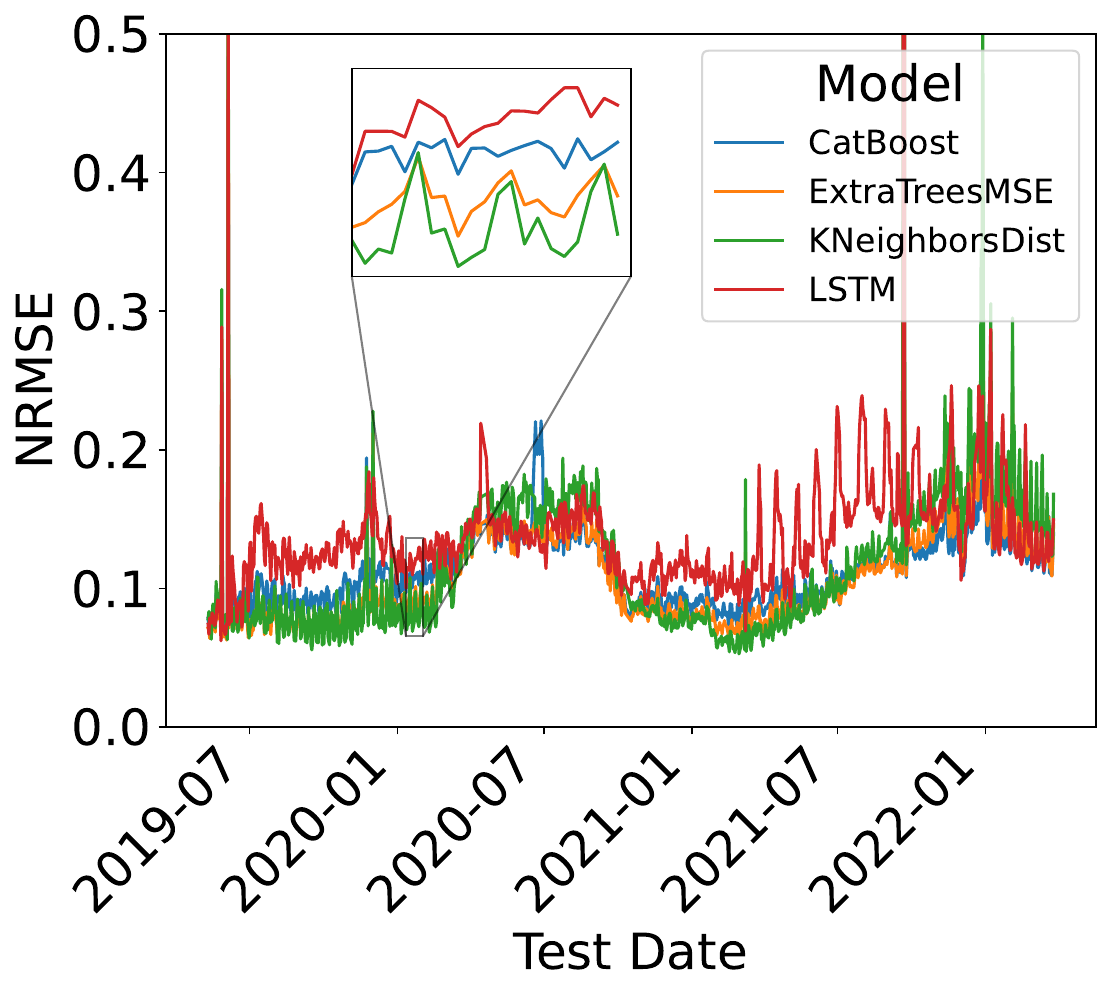}%
  }
  \subfloat[Peak active UEs.\label{fig:peakuser}]{%
    \includegraphics[width=0.235\textwidth]{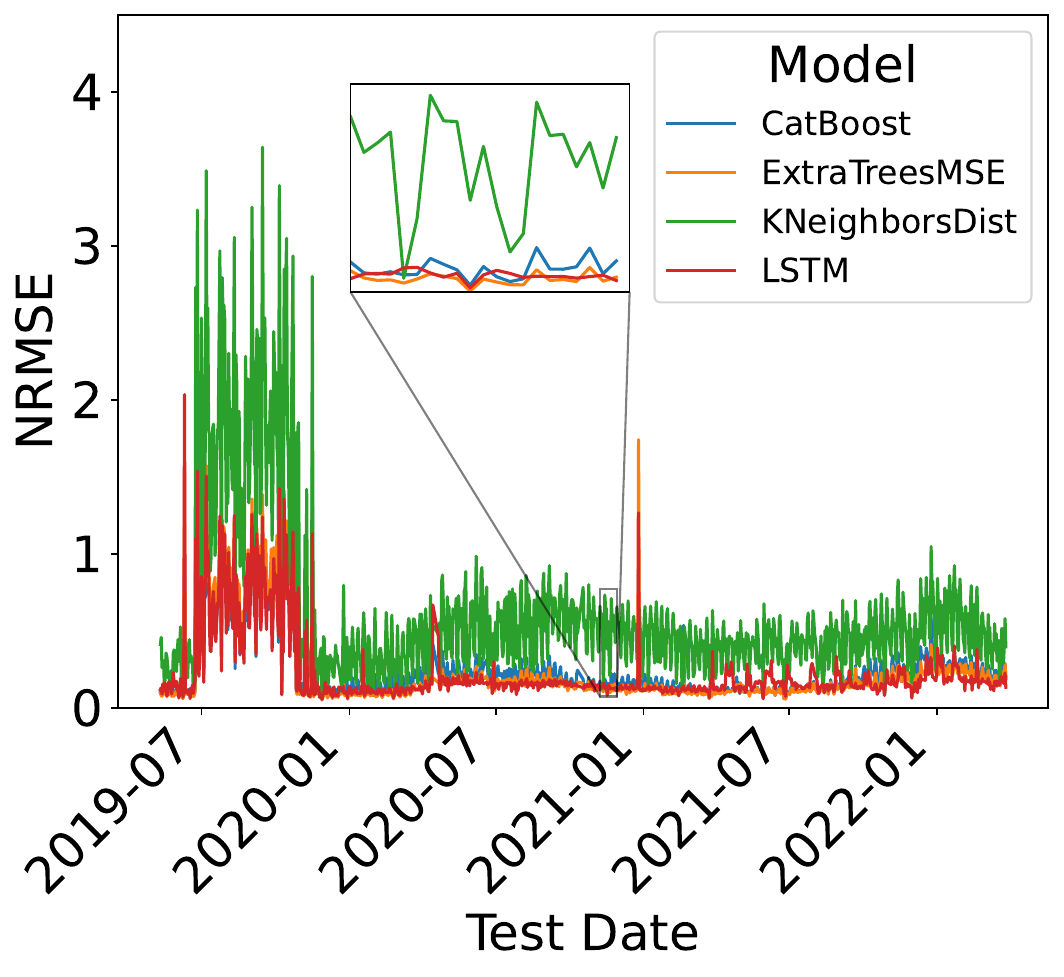}%
  }\\
  \subfloat[Throughput.\label{fig:throughput}]{%
    \includegraphics[width=0.245\textwidth]{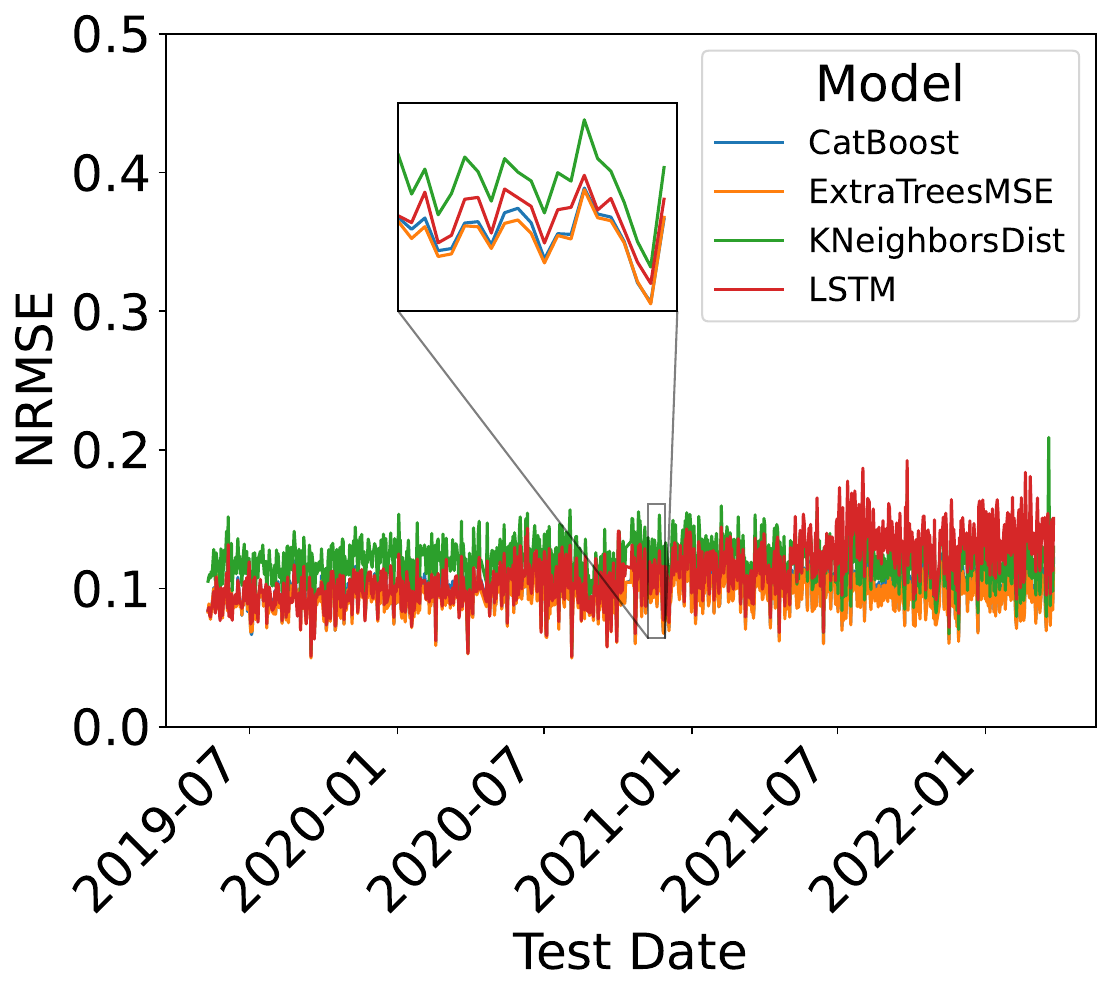}%
  }
  \subfloat[Gap duration ratio.\label{fig:gdr}]{%
      \includegraphics[width=0.245\textwidth]{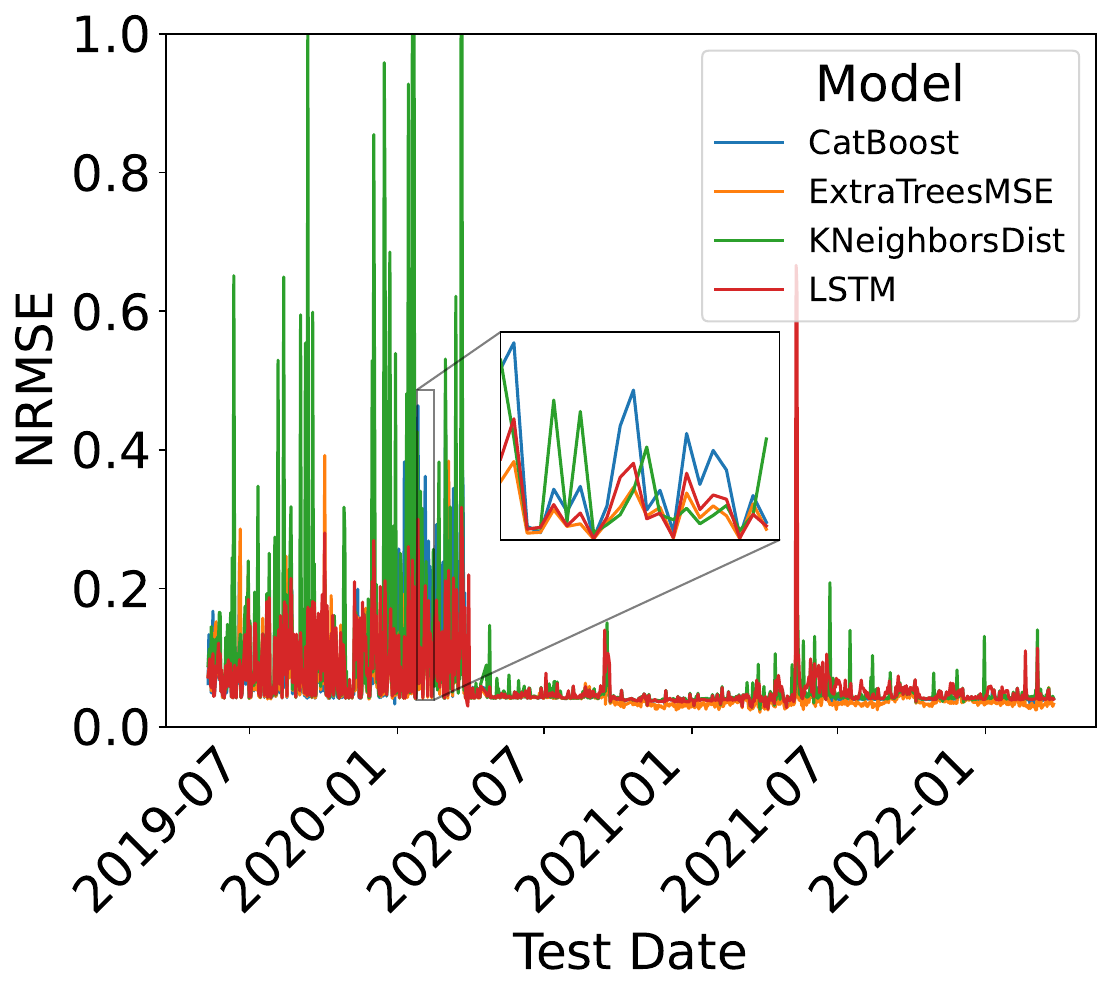}%
  }
  
  \caption{Drift of different models for KPIs of interest. Inset figures 
  exhibit a 3-week view (all starting from Sunday) of NRMSE for the box-selected 
  period. Some data is lost 
  between July, 2019 and January, 2020 for Peak active UEs.
  Note that the y-axes are scaled to different range to accommodate larger errors in
  Fig.\ref{fig:peakuser},~\ref{fig:gdr}.}
  \label{fig:kpi_concept}
\end{figure}

\paragraph{Uncorrelated KPIs exhibit different drift patterns.}
Figure~\ref{fig:kpi_concept} presents the concept drift across time for
three categories of KPIs. Overall, the drift patterns are quite unique for each
class, and they vary in two aspects. First, deviations in NRMSE occur at
different periods of time. For example, Figure~\ref{fig:volume}, NRMSE of downlink volume
experiences a sudden shift in April 2020 (corresponding to the COVID-19 lockdown,
as a unique example of sudden drift), and drifts back to normal values in later months of
2020. Starting from March 2021, the NRMSE gradually increases and peaks around 
January 2022. For the prediction of peak 
users, Figure~\ref{fig:peakuser} demonstrates that July 2019 to November
2019 are harder to predict, because of lost data. Moreover, short-lived, abrupt increases 
in error are more frequent than other KPIs, due to the burstiness 
of GDR. Second, the high-frequency components have different patterns for different
KPIs. By using signal processing techniques like STFT, no obvious weekly pattern
is found on NRMSE of CDR and GDR. But other KPIs have weekly patterns, as the 3-week 
insets also show. 

\paragraph{Different KPIs are most effectively predicted using different models.}
As shown in Figure~\ref{fig:kpi_concept}, we can find a model that
performs relatively well for each target KPI. For all the target KPIs, there is
at least one model with NRMSE less than 0.1 (indicating good prediction 
power~\cite{nau2014notes}) for at least one year. For example, even for GDR, 
the KPI that is most challenging to accurately predict (due to high coefficient 
of variances / dispersion), the average NRMSE from ExtraTrees is 0.055. And 
CatBoost in volume prediction has an average NRMSE of 0.116, with below 0.1 on 
473 days.

\subsection{Analyzing Drift for a Single KPI}\label{subsec:single}

In this section, we analyze how changes in the model used, and how it is trained, 
impact concept drift. We use a single KPI (downlink volume) as an illustrative 
example.

\paragraph{Individual KPI predictions drift consistently across different models.}
As mentioned earlier, we use four different well-trained models to check for 
concept drift over time, while predicting KPIs 180 days in the future. We find (Figure \ref{fig:volume}) 
that all models exhibit the same pattern. For example, between April and October 
2020, all models exhibit a drastic rise in NRMSE due to the COVID-19 pandemic. 
Such consistency is also found in the gradual increase in NRMSE after March 2021. 
We find similar model drift across other metrics such as $R^2$, mean absolute 
error, etc. This pattern holds for each individual KPI. Given our observations 
that models perform similarly well, we use CatBoost for the rest of this paper 
to simplify the presentation. 

\begin{figure}[t]
  \centering
  \subfloat[Training set size. \label{fig:size}]{%
    \includegraphics[width=0.245\textwidth]{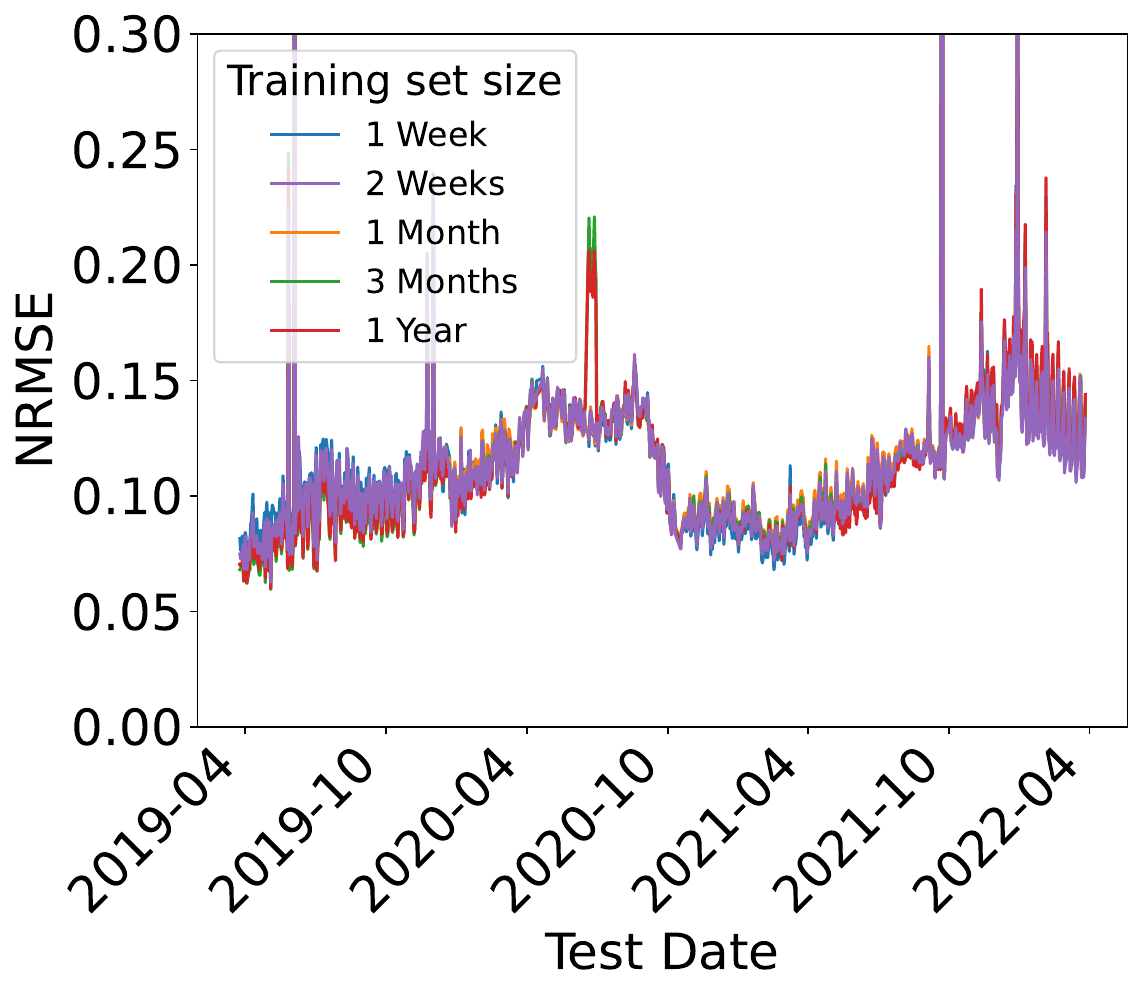}%
  }
  \subfloat[Training set period (training set size = 2 weeks).  \label{fig:period}]{%
    \includegraphics[width=0.235\textwidth]{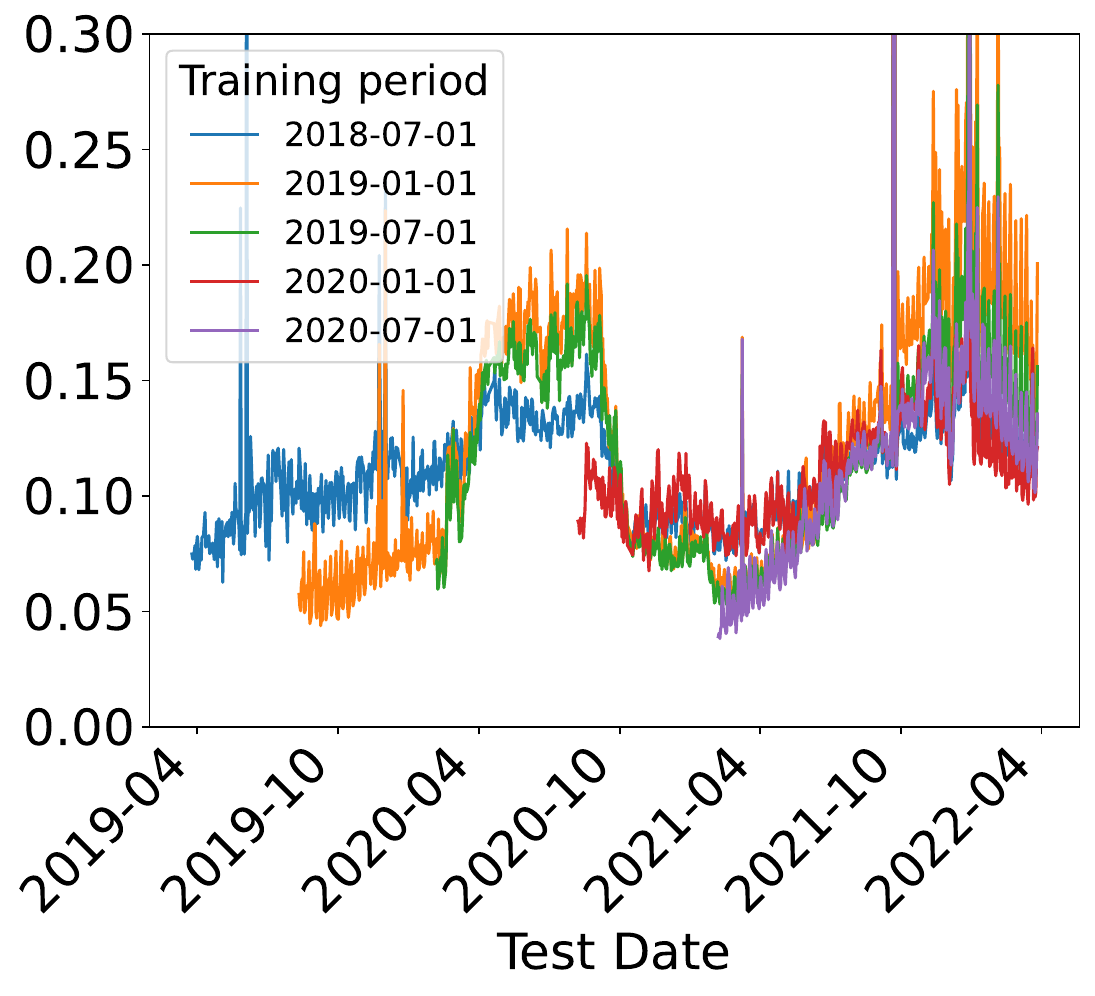}%
  }
  
  \caption{Effects of training set size and training set period on concept drift for the downlink volume KPI}
  \label{fig:parameter} 
\end{figure}

\paragraph{Consistent drift appears when trained using different training set sizes.}
To investigate how parameters of a model could impact concept drift, we
evaluate the impact of varying the training set size. We vary the size of historical data
from eNodeBs with an end date of July 1st, 2018, and retrain models for each
size training window. As Figure~\ref{fig:size} illustrates, all NRMSE values of 
different training set window sizes drift similarly over time. While training 
set windows of one week of data lead to a slightly higher NRMSE, and three months 
and one year witness a glitch around June 2020, the signal pattern remains the same: 
no matter what training set size, NRMSE experiences a sudden increase after 
COVID-19 lockdown and gradually recovers after October 2020, then increases again 
from March 2021 and peaks at January 2022.

In Figure~\ref{fig:size}, we see that the model effectiveness of two weeks 
performs very similarly to that of one year, while the model training time on two 
weeks is 18x more efficient than one year. Given these results, we use 
two weeks for the training set window for the rest of the paper if not otherwise 
specified. This optimizes the tradeoff between performance over time
(\ie, robustness) and efficiency. 

\subsection{Na\"ive Retraining in Practice}\label{subsec:naive_retrain}

In operational networks, a common approach to counteract potential concept drift
is to retrain models regularly. Retraining using the latest data is often considered an effective way to 
deal with concept drift. Many existing solutions~\cite{kantchelian2013approaches,thomas2011design} 
adopt this approach, which outperforms recent dedicated drift mitigation 
methods~\cite{mallick2022matchmaker, you2021learning}. To understand the 
effectiveness of this approach for the \datasetlarge, we retrain a
number of different models using different retraining frequencies. Note that we 
do not fine-tune the models because fine-tuning is very costly for retraining and thus impractical,
due to the need for manual adjustments from human experts. 
For this experiment, we use a training set of 14 days of data to forecast traffic 
volume 180 days in the future, using the CatBoost model. Given a retrain frequency 
$N$, a model is retrained using the latest 14-day data. It is evaluated 
using the NRMSE for the next $N$ days and is then replaced every $N$ days. 

\begin{table}[t!]
  \centering
  \footnotesize
  \newcolumntype{R}{>{\raggedleft\arraybackslash}X}
  \begin{tabularx}{\columnwidth}{cRRRRrrR}
  \toprule
  \textbf{Retraining} & \multicolumn{6}{c}{\textbf{$\Delta \overline{NRMSE}$ of Target KPIs}}  & \multirow{2}{*}{\textbf{\#Retrains}}\\ 
  \cmidrule(ll){2-3} \cmidrule(ll){4-5} \cmidrule(ll){6-7}
  \textbf{Period} & \multicolumn{1}{c}{DVol} & \multicolumn{1}{c}{PU} & \multicolumn{1}{c}{DTP} & \multicolumn{1}{c}{REst}  & \multicolumn{1}{c}{CDR}  & \multicolumn{1}{c}{GDR}  & \\ \midrule
  Static & $-$ & $-$ & $-$ & $-$ & $-$ & $-$ & 0 \\
  7 days & $-40.34\%$ & $-55.36\%$ & $-27.21\%$ & $-48.00\%$ & $47.79\%$ & $-0.38\%$ & 169 \\
  30 days & $-30.66\%$ & $-43.73\%$ & $-21.40\%$ & $-40.12\%$ & $-0.75\%$ & $2.75\%$ & 39 \\
  90 days & $-16.83\%$ & $-16.12\%$ & $-19.07\%$ & $-27.33\%$ & $7.89\%$ & $42.24\%$ & 13 \\
  180 days & $-12.22\%$ & $-0.34\%$ & $-14.85\%$ & $-18.82\%$ & $-4.20\%$ & $76.28\%$ & 6 \\
  365 days & $-2.27\%$ & $-5.13\%$ & $-10.65\%$ & $-11.53\%$ & $-5.97\%$ & $6.07\%$ & 3 \\ \bottomrule
  \end{tabularx}
  \caption{Changes of average NRMSE and number of retrains, over time, for different periodic retraining 
  strategies. }
  \vspace{-10pt}
  \label{tab:retraining}
\end{table}



Somewhat counter to conventional practice, we find that simply retraining the 
model at regular intervals is insufficient for efficiently combating concept drift for a 
diverse, longitudinal dataset. Na\"ive 
retraining is either less effective, or is effective but inefficient, requiring 
frequent retraining. In Table~\ref{tab:retraining}, we show the average NRMSE 
changes compared to the static model. For lower variance KPIs, \ie, DVol, REst, 
and DTP, the more frequently a model is retrained, the better results we obtain. 
Unfortunately, while a 7-day retraining period can mitigate a large portion of 
errors, it is very costly for the network operator. Further, we 
observe that this trend breaks for bursty, high variance KPIs such as CDR (47.79\% 
of error increased every 7 days) and GDR (retrain every 365 days is better than 
every 90 days). In this case, frequent retraining that just na\"ively uses 
all of the most recent data can even adversely affect model performances. 
In Figure~\ref{fig:period}, we train on different 14-day windows of historical 
data from all eNodeBs to verify the pattern. It suggests that, \textit{models 
trained on more recent time periods do not necessarily result in better 
model performance}, because the suddenly changed distribution is more similar
to previous samples. 



Intuitively, a key reason that na\"{\i}ve retraining may not work is that it is triggered 
at regular intervals, even though drift occurrences 
are irregular. It does not take into account when, where, and why drift is occurring.
Thus, at times retraining is not necessary, or it is planned before 
drift actually occurs. This strategy ignores the fact that models trained on more 
recent periods do not necessarily result in better performance 
(Section~\ref{subsec:single}). Further, complete data replacement neglects 
finer-grained error information across samples, throwing away useful samples 
from the past.

Overall, we conclude that while retraining is essential, na\"{\i}vely performing 
it at regular intervals is not sufficient for efficient and explainable drift 
mitigation. It works best at high retraining frequencies and requires specific 
tests across different KPIs to at best fine-tune its performance. Both are 
challenging when run at scale in operational networks. Also, na\"{\i}ve 
retraining neglects finer-grained temporal error information across samples 
and thus loses explainability. This motivates our development of the LEAF 
framework in the rest of this paper.

%% file: sections/leaf.tex
\section{Local Error Approximation of Features (LEAF)}\label{sec:leaf} 

In this section, we introduce \emph{LEAF}, a framework for drift detection,
explanation, and mitigation. We seek to leverage explainable AI methods to
provide us with insights about drift to mitigate it effectively. We describe
first the core intuition behind the LEAF framework, followed by the technical
details for each component.

\subsection{LEAF Framework Overview}\label{subsec:framework} 

LEAF aims to work with any supervised regressors and provide explanations for
black-box models. Based on the explanations provided, targeted mitigation
strategies are applied to compensate for concept drift. Existing solutions face
two major limitations: (1) previous concept drift explanations are limited to
classification problems (\eg,~\cite{jordaney2017transcend,yang2021cade}); (2)
concept drift mitigation is often coarse-grained, only focused on the global
performance metrics (\eg,~~\cite{bach2008paired}).

LEAF overcomes these limitations by implementing a pipeline of three sequential
components: (1) a drift \textit{detector}; (2) a set of tools to
\textit{explain} drift for features; and (3) a drift \textit{mitigator}. Note
that LEAF does not require the use of any specific model nor internal access to
the employed model. Instead, it solely requires access to the previously used
training set data, new data as it arrives, and the generated model. 

\begin{figure}[t]
  \centering
  \includegraphics[width=\columnwidth]{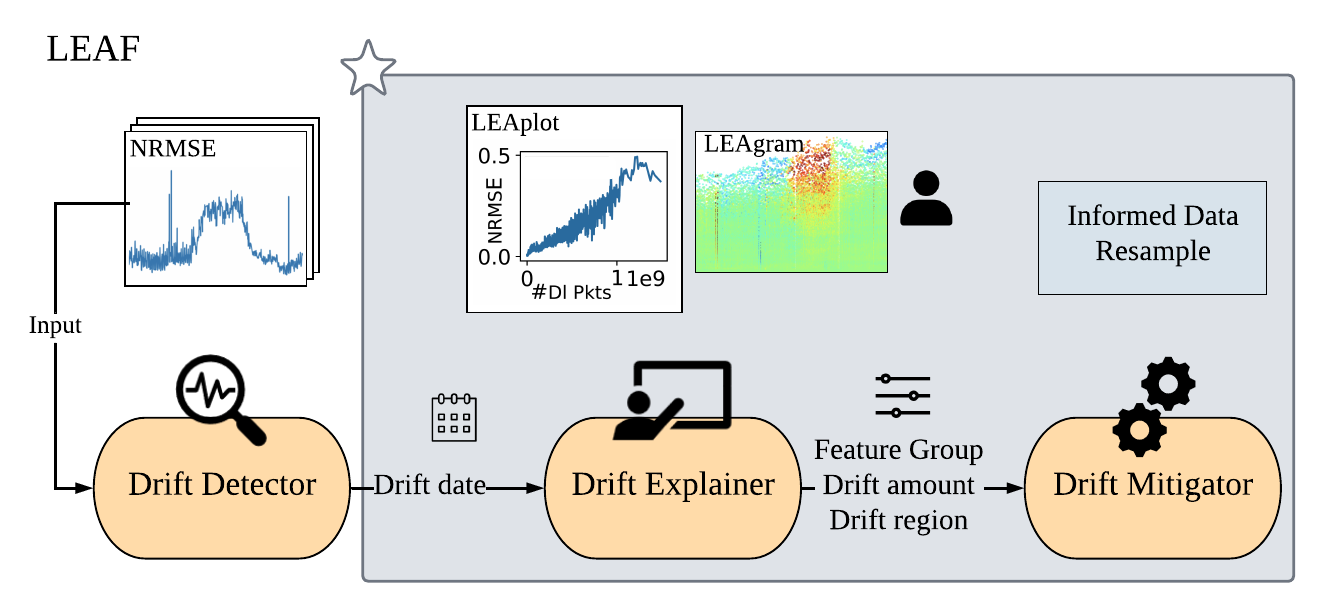}
  \caption{The LEAF framework that detects (Appendix~\ref{subsec:appedix_detection}), 
  explains (\S~\ref{subsec:explanation}), and mitigates (\S~\ref{subsec:mitigation}) concept drift. Our main contributions are in the gray box marked with a star.}
  \label{fig:framework}
\end{figure}

Figure~\ref{fig:framework} shows the three steps in LEAF's pipeline: First, the
detector ingests the outputs of the in-use model in the form of NRMSE
time-series to determine whether drift is occurring. The detector applies the
well-known Kolmogorov-Smirnov Windowing (KSWIN)\footnote{We apply the most
effective drift detection techniques in this well-explored
area~\cite{lu2018learning, gonccalves2014comparative, barros2018large,
bifet2007learning}. We also tested ADWIN, DDM, HDDM, EDDM, PageHinkley, but
KSWIN was the most effective on our NRMSE series.}
method~\cite{raab2020reactive, togbe2021anomalies} on the time-series to
identify a change in the distribution of the output error, providing an
indicator of whether drift is occurring. Although drift detection is critical,
it is not the main focus of this work, as there is significant prior work on
drift detection. A detailed discussion of drift detection is deferred to
Appendix~\ref{subsec:appedix_detection} as the main research contributions of
this work are in the areas of drift explanation and informed mitigation. 

Once drift is detected, the explainer is triggered, indicating the time instance
at which drift occurred. We design LEAF's explainer around the goal of
characterizing errors of a regression model simply based on the model input and
output. We take inspiration from model-agnostic methods (\eg, Partial Dependence
Plot~\cite{friedman2001greedy} and Accumulated Local Effects
Plot~\cite{apley2020visualizing, arzani2021interpretable}) to identify, as well
as visualize, the effect that different features have on prediction errors of
black-box regression models. Doing so, the explainer determines the most
representative features that contribute to drift, and then uses Local Error
Approximation (LEA) to characterize the drift, and offer guidance for model
compensation. 

Based on the error distribution and statistical patterns, the mitigator
\textit{automatically} forgets previous data, and performs informed replacement
by sampling/over-sampling targeted regions. Our insight is that while the global
error metrics provide a good measure of the performance over time, \textit{the
distribution of local errors across samples on each given time instance may be
uneven.} Using this intuition, LEAF's mitigator better compensates for occurring
drift.

The following subsections describe in detail the explainer (\cref{subsec:explanation}) and mitigator (\cref{subsec:mitigation}) modules.


\subsection{Explaining Drift}\label{subsec:explanation} 

The LEAF explanation module is based on the idea of local error decomposition: 
in any regression-based black-box model, local errors may occur in a specific 
range of a given feature or correlated feature set. Once drift is discovered 
in the NRMSE time-series, LEAF determines the extent of error present in 
specific regions of the feature space. To assist mitigation and help operators 
understand drift, we define our goals of the drift explainer as follows: 
(1)~to find and group features that contribute the most to drift, (2)~to 
understand the extent of drift in a given range of values for feature(s), 
(3)~to visualize errors in operational costs and decompose them 
in a spatio-temporal manner. 

\paragraph{Multi-group correlated features.}\label{par:multigroup} Natural
correlations of features are often part of a dataset with a large number of
features~\cite{tolocsi2011classification, gregorutti2017correlation}. 
These correlated features contribute to the
performance of a model simultaneously. LEAF is designed to explain dynamic
feature attributions by finding the representative features that provide the
most valuable error approximations. To achieve this, we first rank features by
permutation-based feature importance (\ie, sensitivity score to
permutation)~\cite{breiman2001random}.
Then, we group features by their correlations. The grouping stops when the
feature has no importance value. Lastly, we choose the most representative
(\ie, highest importance score) feature from each group, as they can represent
the statistical patterns (as well as error distribution) of the group. Using
this technique, we are able to obtain a set of representative features, which
indicates possible factors that lead to drift.

\paragraph{Local error approximation (LEA).} To decompose the error of a model
on any corresponding dataset, we use the selected representative features as
features to inspect. For each feature, we group samples based on the value of
the representative feature into $N$ bins (\ie, quantiles). The higher $N$ is,
the finer the granularity of local errors that can be observed. Next, a
specified error metric (NRMSE by default) is computed for samples within each
bin. We aggregate these $N$ measurements into a vector to represent the error
distribution of each local quantile. This technique yields an approximation of
local errors over the range of the most sensitive and representative feature(s)
for a certain model and corresponding dataset. LEA characterizes the extent of
drift and opens up opportunities to mitigate it in a targeted fashion. LEA,
together with the feature grouping, provides the foundation of the LEAF
framework.

\paragraph{LEAplot and LEAgram.} Starting from LEA, we develop LEAplot and
LEAgram to assist LEAF users to visualize local error components, compare them
across different data subsets, and, overall, better understand root which
features most impact the occurrence of model drift. Given the representative 
features and concerned data subset, LEAplot depicts the quantiles of $N$ 
bins vs. local errors computed in each bin. It shows the distribution and 
extent of error on the feature space. LEAplots exhibit different error 
distributions for representative features from different groups (see Figure~\ref{fig:leaplot} in 
Appendix~\ref{subsec:appedix_detection} for examples). 

LEAgram augments LEAplot by incorporating temporal
information along with error information. It shows the error for individual samples in the entire test set, arranged
temporally. The test set is divided by time interval (\eg, date) and assigns
samples from those divided datasets into $N$ bins, based on the quantiles of
the most important feature with regard to drift. When $N$ is greater than or 
equal to the number of samples on each time interval, errors are shown for 
each sample. 

In operational cellular networks, overestimation leads to different outcomes
compared with underestimation when modeling for capacity planning purposes. For
example, overestimation could result in unnecessary infrastructure expenditure,
while underestimation can lead to user dissatisfaction as infrastructure is not
augmented when it should be. Given these practical issues, we improve the
NRMSE-based approach to create LEAgrams with two modifications: first, we
expand the choice of bin $N$ to be always greater than the number of samples to
see individual sample effects; second, we change the metric to Normalized Error
(NE) to preserve the sign. Example of LEAgram and how it helps interpret drift can be found in the case study 
from Section~\ref{sec:case}.

\subsection{Informed Mitigation}\label{subsec:mitigation} 

The mitigation module is based on the idea of informed adaptation. The key is to
find the optimal data subset to retrain the model and mitigate drift. Given
the information from LEAF's drift explainer, we can understand the features that
contribute most to drift, the amount of drift at each range of values, and the
over/underestimation status of each sample across time. We develop forgetting
and over-sampling strategies using the information, and combine them organically
based on dispersion (\ie, coefficient of variance or Std/Mean in
Table~\ref{tab:target_KPIs}) of the target KPIs.

\paragraph{Forgetting and over-sampling.} When drift is detected, the latest
samples are provided to the explanation module to derive the distribution of
errors from LEA along with the values of the most important feature(s). For example,  
errors (NRMSE) are computed for $N$ bins of the feature range, and they are
associated with each bin. This error distribution $E_L$ shows the area of the
latest dataset that is inaccurately predicted or under-training. 
We first develop a strategy to forget. Because it is a continuous process, 
we apply this method to the last training set. We assign
weights to each sample based on the bin separations of $E_L$ that it
falls into. So that the weight distribution $E_p$ is proportional to the area
of the error. In those areas with high error contributions, we throw away the 
samples with low weights. Then we introduce over-sampling into the pipeline 
from the existing collected dataset (including the latest drifting samples) 
based on weights provided by $E_L$. Afterward, retrains are triggered
using this new training set. Moreover, the detection can
be activated multiple times, so each round of forgetting and over-sampling is
based on the previous round of the restructured training set.

\paragraph{KPI dispersion.} The dispersion of a KPI largely determines the
appropriate mitigation approach and its aggression. If the dispersion of a KPI
around the mean is high, it is more likely to be predicted wrong, and the
forgetting and over-sampling strategy needs to be more aggressive. For example,
for the target KPIs exhibiting a dispersion higher than 1 (\eg, PU,
CDR, GDR), we forget the samples of the original dataset
with linear weights assigned in $E_p$. In terms of over-sampling, we explicitly
use non-linear (cubic) weights~\cite{lahiri2003resampling} of $E_L$ to focus on
the regions of high error in the latest drifting instances. However, if the
dispersion of a target KPI is low (\eg, DVol, DTP, REst), it does not require a
focused over-sampling because of a denser and more even feature space. We
forget the samples of the original dataset with over $95\%$ error, and use
linearized weights of $E_L$ to over-sample the latest drifting instances. Note
that these thresholds and sample fractions are optimal for our dataset and might
need tuning for other datasets, depending on their characteristics.

%% file: sections/casestudy.tex
\section{Case Study on Drift Explanation}\label{sec:case}
\begin{figure*}[th!]
    \centering
    \begin{minipage}{0.3\textwidth}
      \centering
      \includegraphics[width=\textwidth]{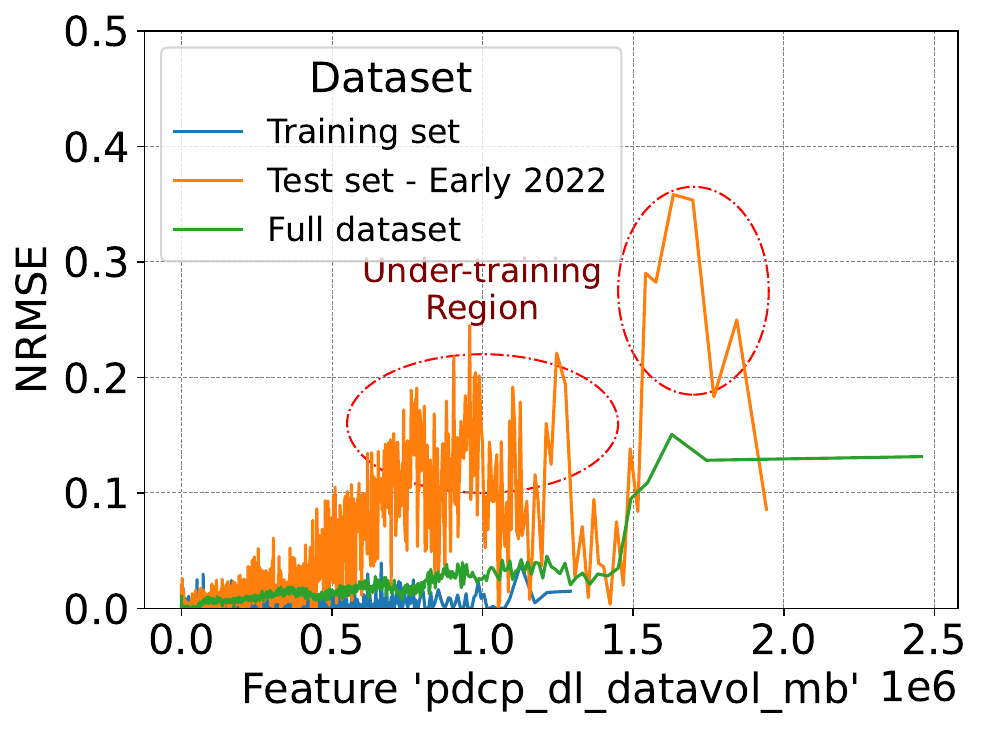}
      \caption{The LEAplots (1,000 bins) that decomposes CatBoost NRMSE time-series. The distribution of estimated local error is shown along with
      the values of the most representative features \code{pdcp\_dl\_datavol\_mb} 
      (downlink volume in Mbps, Group 1).}
      \label{fig:lea_group1}
    \end{minipage}\hfill
    \begin{minipage}{0.685\textwidth}
      \centering
      \subfloat[LEAgram (before mitigation).\label{fig:directed_leagram}]{%
        \includegraphics[width=0.442\textwidth]{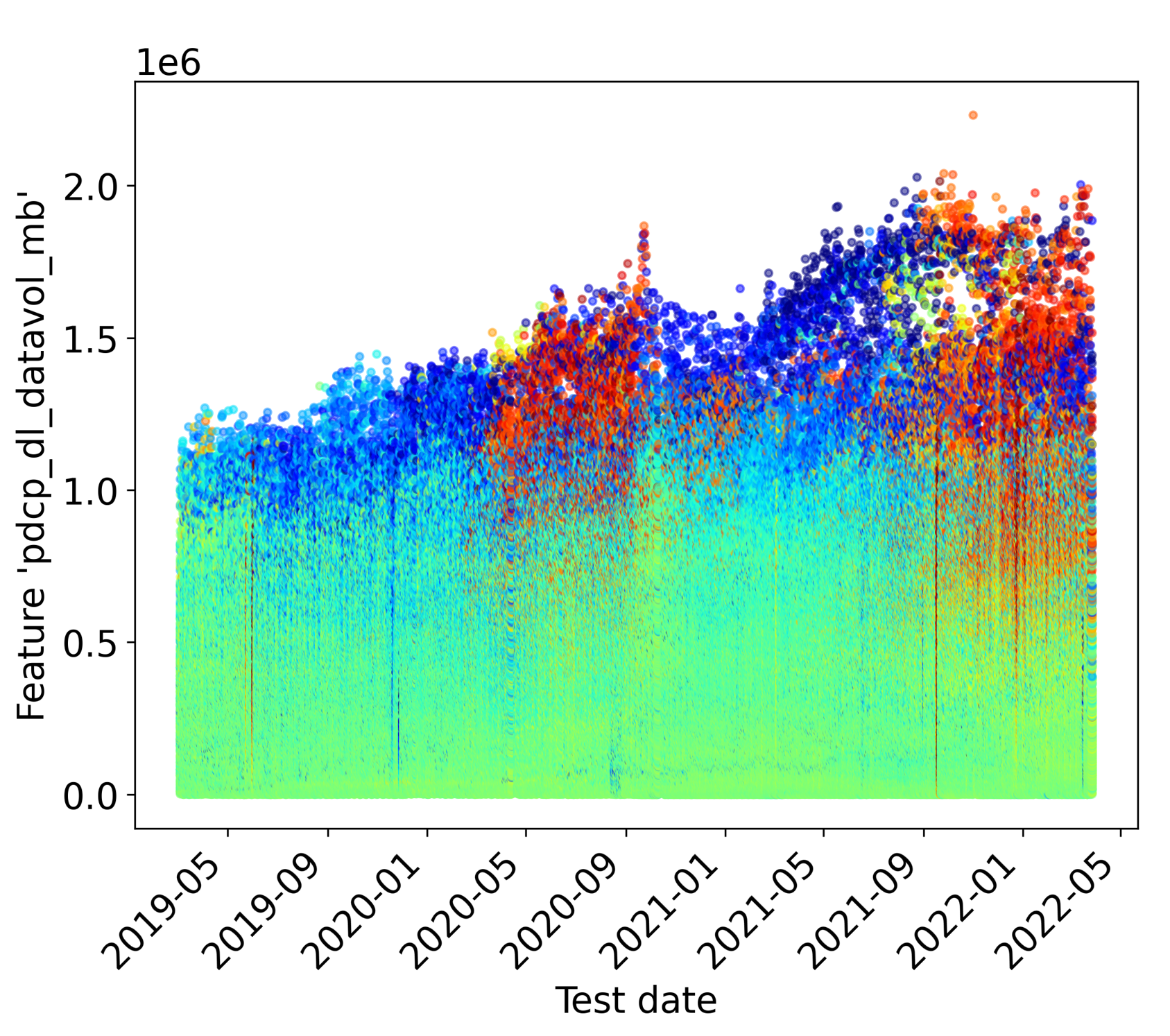}%
      }
      \subfloat[LEAgram (after mitigation).\label{fig:directed_leagram_mitigated}]{%
        \includegraphics[width=0.538\textwidth]{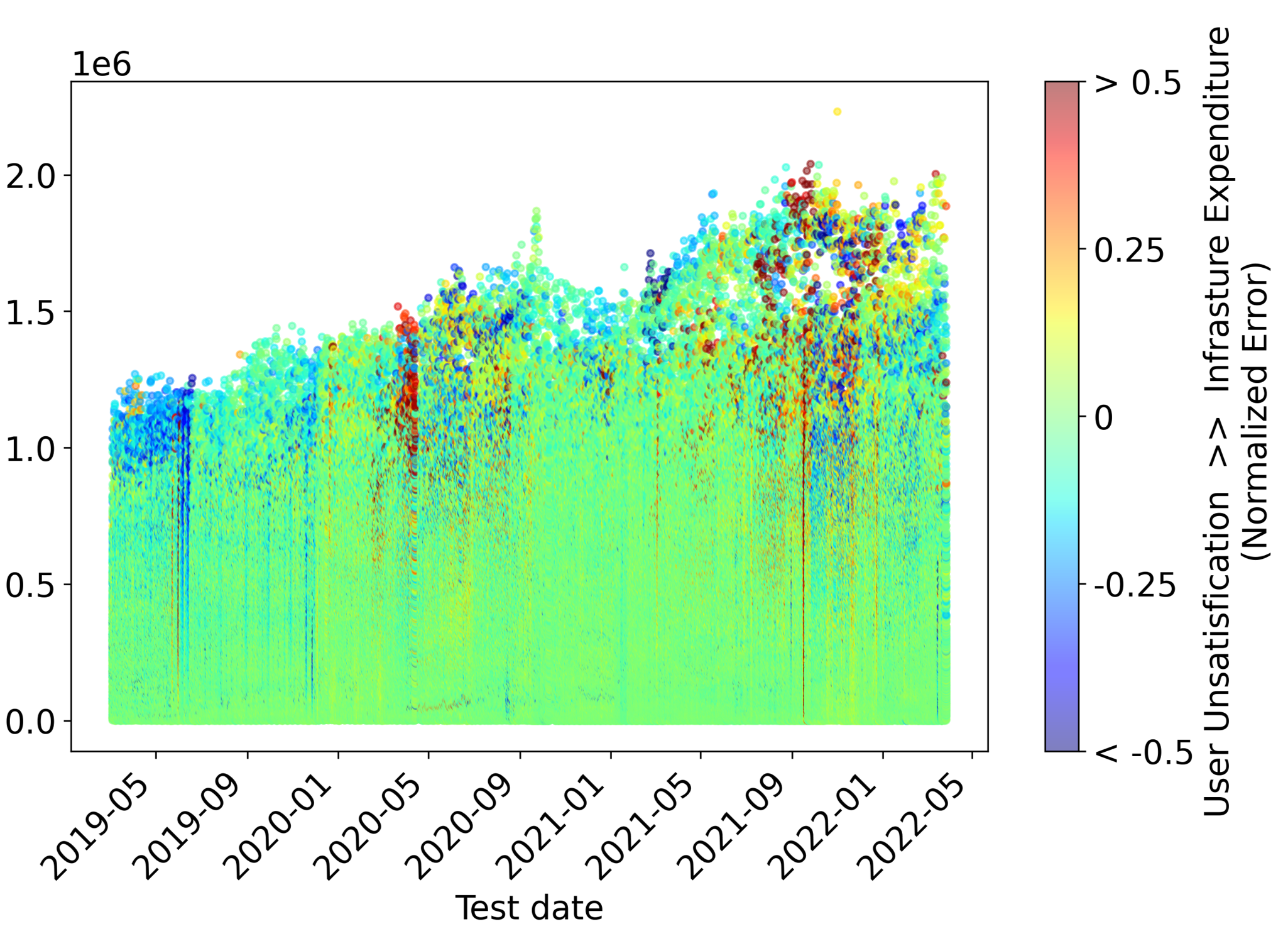}%
      }
      \caption{The LEAgrams that decompose NRMSE time-series, where over- 
      and under-estimation of models are different concerns of operators and 
      they map back to extra infrastructure expenditure and lack of user satisfaction.
      (a) illustrates the decomposition of CatBoost NRMSE. (b) shows errors of 
      mitigated CatBoost model if LEAF applied.}
      \label{fig:directed_leagrams}
  \end{minipage}
  \end{figure*}
In this section, we demonstrate how explanation tools of LEAF help operators to understand
(1) what happens when a single instance of drift occurs (contributing factors and LEAplot); 
(2) an end-to-end interpretation of a model or a mitigation scheme (LEAgram). Here, we showcase
the above by applying \emph{CatBoost to downlink volume forecasting}.

\paragraph{Contributing factors to drift.} In early 2022, our drift 
detector alerts an instance of drift. It triggers the analysis of contributing 
features to understand what leads to drift (to verify if the drift explaining 
features carry the correct semantic meanings, we manually examine with operators). 
Each representative feature indicates a possible factor that leads to drift. The 
most representative feature of the 1st group is \textbf{\code{pdcp\_dl\_datavol\_mb}}, 
the history of downlink volume itself, which serves as a sanity check. 
Highly correlated features include different 
stages of traffic communication: side channels that carry or reflect volume 
(\eg, average downlink packets from UE), connection establishments and releases (\eg,
establishments of RRC), and etc. All 32 features
from this group contribute to error jointly, and thus will be used to explain
the drift as a whole. Representative features from other groups provide
different views of possible reasons. The 2nd group contains features
that reflect another reason: \textbf{\code{badcoveragemeasurements}}, which measures the
bad coverage of eNodeBs in percentage. This often raises from the geographical distance
between an eNodeB and a UE, or destructive interference, which influences the
downlink volume. The 3rd representative feature is
\textbf{\code{rtp\_gap\_ratio\_medium\%}}, which is related to specifically losing voice
data packets during VoLTE call. With it being zero in most cases, bursty gaps between packets indeed affect
traffic delivery, hence downlink volume. 

\paragraph{Local error illustrations of drift using LEAplot.} 
Once the representative features are found, we apply LEA on them and illustrate 
the locally decomposed error of the drift by LEAplot per feature group, for 
a better understanding of the extent of drift within different value ranges of the feature.
Operators can also localize the eNodeBs with bad performance.
Figure~\ref{fig:lea_group1} shows the LEAplot of the top representative feature.
\code{pdcp\_dl\_datavol\_mb} (the history of the downlink volume in Mb) is the most 
representative feature from Group 1
for this model. Although the training set has a very low error when the feature
value changes, when its value is between 0.6e6 and 1.3e6, errors in the ``Early 2022'' test
set are more than 10x those of the training set and nearly 6x of those in the
full test set. For the value above 1.5e6, the training set does not cover the range 
so the test sets observe high errors on it. 
Note that the LEAplots of different representative features exhibit different 
distributions (see Appendix~\ref{subsec:appedix_detection}), which can be used 
for multi-group mitigations. From LEAplot, we further trace back the eNodeBs that
exhibit high errors. The top 5\% of error mostly comes from eNodeBs located
at suburban areas, because users there change their mobility pattern around 
this instance of drift, which is after the winter break.

\paragraph{End-to-end interpretation of performance using LEAgram.} 
LEAgram assists to derive the explanation, by 
revealing the changes of local errors and their durations over time. 
Operators can associate spatio-temporal changes of error with plausible 
reasons (\eg, hardware installations, firmware updates, etc.) based on the location and duration of events.
For this case, we interpret the error direction and extent before and after mitigations. 
Figure~\ref{fig:directed_leagram} shows the LEAgram. From March 15, 2020 to November 
1, 2020, when the value of feature \code{pdcp\_dl\_datavol\_mb} is above 1e6 Mb, 
large positive errors occur, implying overestimation. It indicates possible reasons
as operators can correlate them with less user mobility, because people switch to broadband networks rather than 
cellular networks during the lockdown, thus less actual demands. Overestimations from this 
model occur again after October 2021, when the value of feature 
\code{pdcp\_dl\_datavol\_mb} is above 0.6e6. If the operator were to base 
decisions on the output of this model, they may unnecessarily build new 
infrastructure. The model also has negative prediction errors above a value 
of 1e6, which could lead to user dissatisfaction as the operator may predict 
a lower demand than in reality. LEAgram also assists in demonstrating the performance of 
mitigation schemes over time. Figure~\ref{fig:directed_leagram_mitigated} shows the mitigating
effects of LEAF not only during COVID-19 lockdown, but also in early 2022 for 
gradual shifts. Overall, 32.68\% of reduction in
$\Delta \overline{NRMSE}$ is shown compared to Figure~\ref{fig:directed_leagram}, 
with a major mitigation focus on the errors at the tail. 

%% file: sections/evaluation.tex
\section{Evaluation}\label{sec:evaluation} In this section, we evaluate how the
LEAF framework mitigates drift compared to baseline techniques. We present how
different mitigation schemes improve end-to-end model performance over the
\dataset. We initially focus on the \dataset to provide an ``apples
to apples'' comparison across models and to understand whether the inherently 
changing nature of the dataset lowers performance for some models. We then 
evaluate using the \datasetlarge. Unless specified, we present the
performance of LEAF with a single feature group. We focus on the 
comparison across mitigation schemes as Section~\ref{sec:drift} already 
established the need for applying techniques to combat drift. A comparison 
between LEAF and the static model is available in Appendix~\ref{subsec:appedix_e2e}.

\subsection{End-to-End Comparison Across Mitigation
Schemes}\label{subsec:eval_end2end}

In this subsection, we compare the average NRMSE over the duration of the dataset against two
baselines: the na\"ive retraining scheme (see
Section~\ref{subsec:naive_retrain}) that retrains the model every 30 or 90 days
(we choose these two frequencies because they present the best performance
versus retraining frequency); and triggered retraining, which only utilizes the
drift detector information, \ie, retrain the model using the latest available
data whenever drift is detected. For LEAF, we also control the number of feature
groups used during the mitigation phase. We show results for one, three, and
five feature groups as they are the best performing configurations. To keep the
evaluation fair across methods, we use the same amount of data for each retrain
operation, which also controls the amount of time needed for a single retrain
across schemes. The mitigation effectiveness is compared against a static model
(trained on 14 days of data before July 1st, 2018) for each target KPI.

\begin{figure}[t]
  \centering
    \subfloat[Downlink Volume]{%
      \includegraphics[width=0.45\columnwidth]{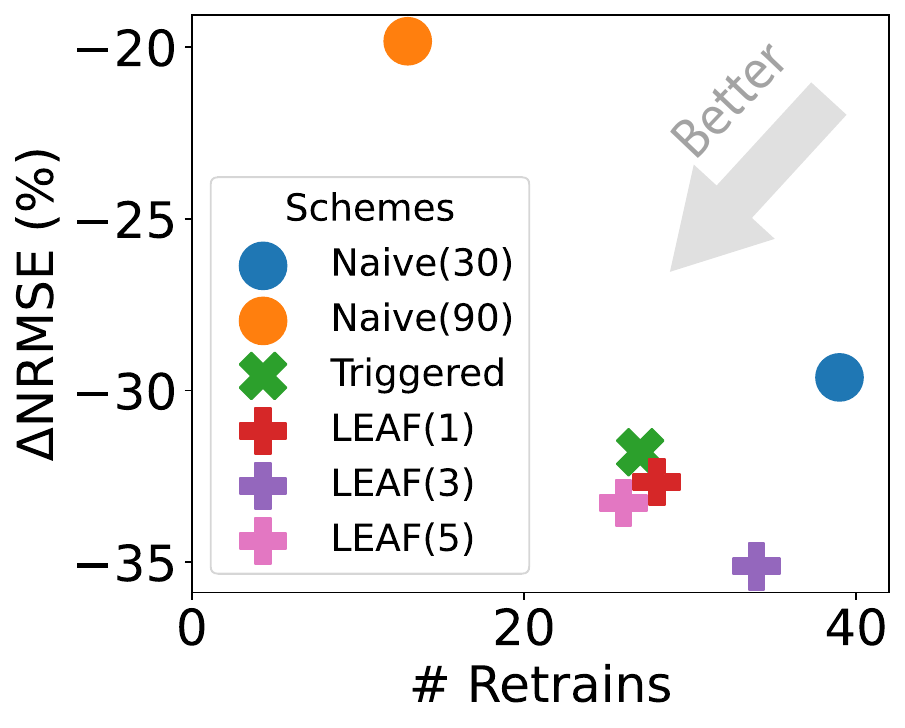}%
      \label{fig:DVol_tradeoff}
    }
    \subfloat[Call drop Rate]{%
      \includegraphics[width=0.44\columnwidth]{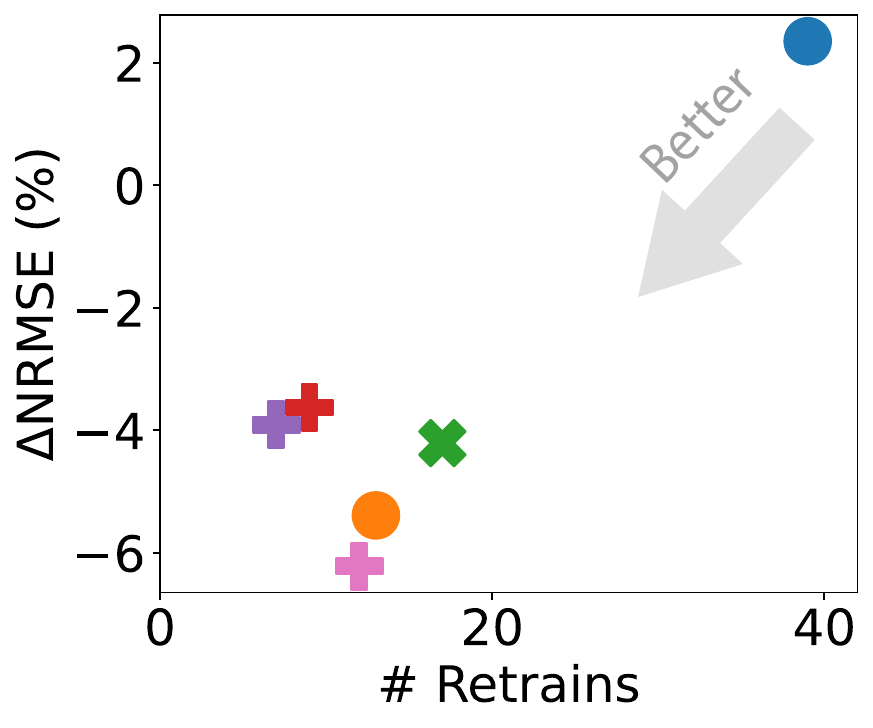}%
      \label{fig:CDR_tradeoff}
    }
    \caption{$\Delta \overline{NRMSE}$ vs. $\#Retrains$ under different 
    mitigation schemes using CatBoost. For na\"ive retraining, 
    number in parentheses denotes the retraining period in days. For LEAF,
    it represents the number of feature groups being used.
    Bottom left has the best mitigation effectiveness and lowest retraining
    cost.}
    \label{fig:e2e_tradeoff}
\end{figure}

Figure~\ref{fig:e2e_tradeoff} shows the trade-off between $\Delta \overline{NRMSE}$
and the number of retrains required by each scheme using CatBoost across
KPIs (we only show two as they present similar patterns, full evaluations are 
in Appendix~\ref{subsec:appedix_full_eval}). Such trade-off provides insights not only on
the performance of each mitigation scheme, but also on its applicability in
practice in an operational network where each retrain operation might come at a
cost. Our goal is to find the scheme that achieves the best mitigation 
effectiveness first, while balancing the alternative goal of few retrains. 
As such, the bottom left of each subfigure represents the best option.

In the figure we observe that, as expected, na\"ive retraining every 30 days
requires the highest number of retrains, while its mitigation effectiveness
never outperforms LEAF. Na\"ive retraining every 90 days, however, requires the
fewest retrains except for CDR mitigation. However, the mitigation
effectiveness is frequently inferior to LEAF's, sitting in the top left corner
in Figure~\ref{fig:DVol_tradeoff}. While this mitigation
scheme seems to perform well for CDR, it does not achieve the best
mitigation. The triggered mitigation scheme often lies in the middle, never
outperforming other schemes on either metric. It also has exponential
errors for KPIs like GDR, making it less practical among the schemes since it
does not guarantee performance improvements for a model after mitigation. 

Finally, LEAF consistently outperforms all baseline schemes. Thanks
to its approach focused on using error explanation and informed mitigation, 
LEAF already exceeds the performance of other methods across
KPIs, except for CDR, even when only using one representative feature.
When considering additional feature groups, it can achieve either higher
effectiveness or a lower number of retrains needed. Compared to triggered
retraining, LEAF is able to mitigate more errors, with a similar or slightly
higher number of retrains (\eg, all but CDR). For CDR, LEAF is
able to mitigate a similar amount of errors with 30.8\% fewer retrains.
Notably, the number of feature groups used in LEAF also impacts mitigation
effectiveness. This will be further discussed in the next subsection.

\subsection{Sensitivity Analysis}\label{subsec:eval_discussion}
In this section, we explore which configurations work best depending 
on the number of feature groups employed in LEAF and its performance across 
models and KPIs.

\paragraph{Different number of feature groups.} 
Multi-group LEAF forgets and over-samples the input dataset according to the
error distributions from more than one representative feature. Although
the amount of data being resampled remains the same, multi-group LEAF
iteratively optimizes which retrain data to use based on the input of the drift
explanation component. As shown in Figure~\ref{fig:e2e_tradeoff}, the amount of
errors being mitigated tends to increase when more feature groups are used,
except for GDR. 0.34\% to 2.83\% more errors are mitigated by multi-group LEAF
compared to single-group LEAF. However, the optimal number of feature groups is
not constant across KPIs. For DVol, PU, and REst, three feature groups lead to
the highest mitigation effectiveness. For DTP and CDR, five feature groups work
best. As the most dispersed KPI, GDR is an outlier that reaches optimality with
only one feature group. Feature importance and the number of features within each
group might be the factors that affect the effectiveness of the mitigation.

\begin{table}[!t]
  \centering
\footnotesize
\begin{tabularx}{0.5\textwidth}{Xcrlrlrl}
  \toprule
  \multirow{2}{*}{\textbf{Model}} & \multirow{2}{*}{\textbf{KPIs}} & \multicolumn{6}{c}{\textbf{$\Delta \overline{NRMSE}$ ($\#$Retrains)  of Mitigation Schemes}}  \\ 
  \cmidrule(llll){3-8}
   &  & \multicolumn{2}{c}{Na\"ive$_{30}$} & \multicolumn{2}{c}{Triggered} & \multicolumn{2}{c}{LEAF} \\ \midrule
   \multirow{6}{*}{\textbf{CatBoost}} & DVol & $-29.62\%$ & (39) & $-31.80\%$ & (27) & \cellcolor[HTML]{b7b7b7}$-32.67\%$ & (28) \\
   & PU & $-44.88\%$ & (39)  & $-35.06\%$ & (25) & \cellcolor[HTML]{b7b7b7}$-46.59\%$ & (35) \\
   & DTP & $-20.02\%$ & (39) &  $-23.84\%$ & (28) & \cellcolor[HTML]{b7b7b7}$-24.30\%$ & (31) \\
   & REst & $-35.41\%$ & (39) & $-38.38\%$ & (25) & \cellcolor[HTML]{b7b7b7}$-38.44\%$ & (31) \\
   & CDR & $2.35\%$ & (39) & \cellcolor[HTML]{b7b7b7}$-4.21\%$ & (17) & $-3.63\%$ & (9) \\
   & GDR & $3.37\%$ & (39) & $44.56\%$ & (17) & \cellcolor[HTML]{b7b7b7}$-6.24\%$ & (19) \\ \midrule
   \multirow{6}{*}{\textbf{ExtraTrees}} & DVol & $-24.77\%$ & (39) & $-28.17\%$ & (32) & \cellcolor[HTML]{b7b7b7}$-30.64\%$ & (32) \\
   & PU & $-44.26\%$ & (39) & \cellcolor[HTML]{b7b7b7}$-50.76\%$ & (26) & $-45.83\%$ & (27) \\
   & DTP & $-18.13\%$ & (39) & $-21.63\%$ & (32) & \cellcolor[HTML]{b7b7b7}$-22.59\%$ & (23) \\
   & REst & $-31.95\%$ & (39) & $-34.29\%$ & (22) & \cellcolor[HTML]{b7b7b7}$-36.13\%$ & (29) \\
   & CDR & $2.10\%$ & (39) & $8.08\%$ & (20) & \cellcolor[HTML]{b7b7b7}$-0.20\%$ & (11) \\
   & GDR & $-0.58\%$ & (39) & $33.67\%$ & (17) & \cellcolor[HTML]{b7b7b7}$-14.26\%$ & (19) \\ \midrule
   \multirow{6}{*}{\textbf{LSTM}} & DVol & \cellcolor[HTML]{b7b7b7}$0.54\%$ & (39) & $14.12\%$ & (21) & $2.67\%$ & (19) \\
   & PU & $37.11\%$ & (39) & $3.76\%$ & (25) & \cellcolor[HTML]{b7b7b7}$-20.48\%$ & (18) \\
   & DTP & $17.08\%$ & (39) & $-0.82\%$ & (14) & \cellcolor[HTML]{b7b7b7}$-37.13\%$ & (20) \\
   & REst & $6.78\%$ & (39) & $5.78\%$ & (27) & \cellcolor[HTML]{b7b7b7}$4.21\%$ & (26) \\
   & CDR & $-33.22\%$ & (39) & $-21.39\%$ & (10) & \cellcolor[HTML]{b7b7b7}$-71.52\%$ & (11) \\
   & GDR & $0.41\%$ & (39) & $-8.58\%$ & (14) & \cellcolor[HTML]{b7b7b7}$-16.29\%$ & (13) \\ \midrule
   \multirow{6}{*}{\textbf{KNeighbors}} & DVol & \cellcolor[HTML]{b7b7b7}$-8.26\%$ & (39) & $-4.11\%$ & (16) & $-4.47\%$ & (24) \\
   & PU & $-34.09\%$ & (39) & \cellcolor[HTML]{b7b7b7}$-37.99\%$ & (16) & $-18.11\%$ & (20) \\
   & DTP & \cellcolor[HTML]{b7b7b7}$-4.73\%$ & (39) & $-4.03\%$ & (18) & $-1.53\%$ & (22) \\
   & REst & \cellcolor[HTML]{b7b7b7}$-26.69\%$ & (39) &  $-25.86\%$ & (25) & $-22.10\%$ & (16) \\
   & CDR & $9.44\%$ & (39) & $7.35\%$ & (11) & \cellcolor[HTML]{b7b7b7}$4.69\%$ & (12) \\
   & GDR & $-8.13\%$ & (39) & \cellcolor[HTML]{b7b7b7}$-23.40\%$ & (19) & $-6.12\%$ & (13) \\ \bottomrule
  \end{tabularx}
  \caption{Effectiveness of mitigation schemes measured in $\Delta \overline{NRMSE}$ and $\#Retrains$ 
  (both are the lower the better) using \dataset. We include 
  models from different model families over a variety of KPIs. The scheme with the 
  highest performance is shaded in gray.}
  \label{tab:delta_nrmse}
\end{table}
\paragraph{Different models.}
Table~\ref{tab:delta_nrmse} presents a summary of the $\Delta \overline{NRMSE}$
across different forecasting models (the lower, the better). We highlight in
gray the best-performing scheme for each model and KPI. As expected, we find
that each model responds differently to the mitigation schemes. LEAF outperforms
baselines across most combinations of models and KPIs, with the exception of
KNeighbors. For CatBoost and ExtraTrees models, LEAF is either the most
effective or very close to the best performing scheme across all KPIs. Further,
LEAF consistently mitigates drift across all models, \ie, their $\Delta
\overline{NRMSE}$s are always negative. Other schemes, like na\"ive
retraining or triggered retraining, do not guarantee model improvement across
KPIs. For example, for CDR and GDR, both schemes end increasing errors after
mitigation up to 44.56\% in comparison to the static models. 

Additionally to improving model performance, LEAF achieves the best results
while requiring 10.3\% to 76.9\% fewer retrains for CatBoost and 17\% to
71.8\% fewer for ExtraTrees when compared to na\"ive retraining every 30 days.
We validate this pattern by testing on other boosting (LightGBM) and
bagging (Random Forest) algorithms and notice that it holds across these model
families. 

For LSTM, we find that LEAF is drastically better at reducing NRMSE. Despite
achieving slightly worse performance for DVol and REst, LEAF is the most
effective scheme for the other KPIs. Moreover, when effective, LEAF
reduces the NRMSE by a large margin compared with the second best results. By
applying LEAF, LSTM achieves 7.71\% to 50.13\% less NRMSE than triggered
retraining. Surprisingly, CDR errors are reduced by 71.52\% when only 11
retrains are required.

Finally, we observe that LEAF does not mitigate well KNeighbors (and other
distance-based models). We believe this behavior is rooted in the expressiveness 
of distance-based models and how they generalize~\cite{losing2016knn}. KNeighbors 
employs a lazy regressor that memorizes all the training set and uses the 
least distance to the nearest neighbor to perform predictions. While 
over-sampling the error region can show targeted improvement, the originally good 
regions may perform worse because of the unbalanced added samples. In contrast, 
bagging methods train weak learners in parallel~\cite{geurts2006extremely} (or sequentially with
boosting~\cite{dorogush2018catboost}). As individual learners learn new samples 
more independently, they are less affected by previous learners during targeted 
mitigation.

\paragraph{Different KPIs.}
As observed in the previous analysis, we find that different KPI time-series are
mitigated more or less effectively depending on their own characteristics.
For example, in CatBoost and ExtraTrees mitigation, PU has the highest
NMRSE among KPIs across models and it is among the three most challenging KPIs
to mitigate, along with CDR and GDR. This behavior can be intuitively attributed
to the fact that PU has the highest NMRSE values over time caused by sudden data
losses. This generates an inherent high variability and causes model collapse.
CDR and GDR have low baseline NMRSEs but are difficult to mitigate due to the
high-frequency variation they experience. We validate this hypothesis by looking
at the coefficient of variance, where PU, CDR, and GDR have the highest coefficients of variation of $1.34$, $1.35$ and
$2.12$, respectively. 
These KPIs, when mitigated by a method such as triggered retraining, actually
lead to a large increase in NMRSE. Less dispersed KPIs like DVol, DTP, and REst,
although still present drift, are easier to adapt, because of more homogenous
distribution changes.

\subsection{LEAF Effectiveness on Evolving Infrastructure}\label{subsec:eval_transfer}
We present the effectiveness when applying LEAF on models that
aim to forecast KPIs for an evolving infrastructure, \ie, where new eNodeBs are
being constantly deployed. In the previous analysis, we focused on a fixed
number of eNBs in the \dataset to evaluate internal drift factors like software
upgrades and user behavior pattern changes. Here, we use the \datasetlarge to
test the influence of daily sample numbers and changing infrastructure.

\begin{table}[!t]
  \centering
\small
\begin{tabularx}{0.5\textwidth}{Xcrlrlrlrlrl}
  \toprule
  \multirow{2}{*}{} & \multirow{2}{*}{\textbf{KPIs}} & \multicolumn{6}{c}{\textbf{$\Delta \overline{NRMSE}$ ($\#$Retrains) of Mitigation Schemes}}  \\ 
  \cmidrule(llll){3-8}
   &  & \multicolumn{2}{c}{Triggered} & \multicolumn{2}{c}{LEAF} & \multicolumn{2}{c}{LEAF*}\\ \midrule  
   & DVol & $-31.80\%$ & (27) & $-32.67\%$ & (28) & \cellcolor[HTML]{b7b7b7}$-35.12\%$ & (34) \\
   & PU & $-35.06\%$ & (25) & $-46.59\%$ & (35) & \cellcolor[HTML]{b7b7b7}$-47.62\%$ & (27) \\
   \textbf{Fixed} & DTP &  $-23.84\%$ & (28) & $-24.30\%$ & (31) & \cellcolor[HTML]{b7b7b7}$-24.64\%$ & (30) \\
   \textbf{Dataset} & REst & $-38.38\%$ & (25) & $-38.44\%$ & (31) & \cellcolor[HTML]{b7b7b7}$-41.27\%$ & (32)\\
   & CDR & $-4.21\%$ & (17) & $-3.63\%$ & (9) & \cellcolor[HTML]{b7b7b7}$-6.22\%$ & (12)\\
   & GDR & $44.56\%$ & (17) & \cellcolor[HTML]{b7b7b7}$-6.24\%$ & (19) & \cellcolor[HTML]{b7b7b7}$-6.24\%$ & (19) \\ \midrule
       
   & DVol & $-30.76\%$ & (24) & $-32.09\%$ & (37) & \cellcolor[HTML]{b7b7b7}$-32.80\%$ & (30)\\
   & PU & $-50.89\%$ & (24) & $-45.75\%$ & (24) & \cellcolor[HTML]{b7b7b7}$-51.72\%$ & (26)\\
   \textbf{Evolv.} & DTP & $-22.19\%$ & (30) & \cellcolor[HTML]{b7b7b7}$-22.58\%$ & (27) & \cellcolor[HTML]{b7b7b7}$-22.58\%$ & (27)\\
   \textbf{Dataset} & REst & $-44.01\%$ & (25) & $-43.66\%$ & (26) & \cellcolor[HTML]{b7b7b7}$-48.01\%$ & (33) \\
   & CDR & $-6.79\%$ & (7) & $-1.33\%$ & (9) & \cellcolor[HTML]{b7b7b7}$-7.15\%$ & (8) \\
   & GDR & \cellcolor[HTML]{b7b7b7}$-13.21\%$ & (15) & $-2.06\%$ & (13) & $-11.99\%$ & (17)\\ \bottomrule
  \end{tabularx}
  \caption{Effectiveness of different mitigation schemes measured in $\Delta \overline{NRMSE}$ and $\#Retrains$ 
  (both are the lower the better) using both datasets. We show the best scheme of 
  multi-group LEAF and denote it as LEAF*. The table with na\"ive retraining 
  can be found in Appendix~\ref{subsec:appedix_full_eval}.}
  \label{tab:delta_nrmse_evolv}
\end{table}

We evaluate the effectiveness of the mitigation schemes across both datasets and
present results in Table~\ref{tab:delta_nrmse_evolv}. As before, we only show
the best scheme of multi-group LEAF and denote it as LEAF*. Both
na\"ive retraining schemes have very close performance across the majority of
KPIs, with the exception of CDR and GDR with a retraining frequency of 30 days.
We assume that this improvement can be recollected to the high retraining
frequency of the first scheme that enables the model to quickly capture new
eNodeBs as they are deployed. This trend is also observed for triggered
retraining: working on the \datasetlarge greatly improves performance across the
majority of KPIs, confirming that employing a timely drift detector is
beneficial to detect changes in the data when new elements are deployed in the
infrastructure. Finally, we observe that both LEAF and LEAF* performance remains
consistent across datasets, while also remaining the best performing mitigation
strategy. This demonstrates that LEAF outperforms other baselines as it both
integrates a detector to identify when drift is occurring as well as a more
effective mitigation strategy that can use features error information to better
target the data to use for retraining.

%% file: sections/related.tex
\section{Related Work}\label{sec:related} 



\paragraph{Drift in network management.} 
Machine learning has been applied to many networking problems, including
anomaly detection~\cite{shon2007hybrid, shon2005machine}, intrusion
detection~\cite{sinclair1999application, dong2016comparison}, cognitive network
management~\cite{ayoubi2018machine}, and network
forecasting~\cite{mei2020realtime, chinchali2018cellular}. Few
studies have explicitly explored concept drift in the networking context.
It is, however, well-known that network traffic is inherently variable over
time. For example, features such as latency can also drift over
long periods of time~\cite{nikravesh2014mobile}. Moreover, the COVID-19 pandemic
has provided a unique example of sudden drift and its effects on network
traffic patterns, network usage, and resulting model
accuracy~\cite{liu2021characterizing, feldmann2020lockdown,
lutu2020characterization, comcast2020covid, mckeay2020parts}. Sommer and Paxson
articulate that applying machine learning to anomaly detection is fundamentally
difficult in large-scale operational networks~\cite{sommer2010outside} due to
these (and other) challenges. All these
factors contribute to concept drift in predictive models, yet, to our
knowledge, this paper is the first to explore the effects of these types of
changes on models for cellular network.

\paragraph{Drift in other related fields.} Concept
drift has also been studied across other real-world contexts. For example, spam
detection faces drift challenges, as spammers may actively change the
underlying distribution of their messages~\cite{fdez2007applying}. Recommender
systems are another context where drift occurs: user side-effects such as
preferences and item side-effects such as popularity both change over
time~\cite{gama2014survey, lu2015recommender}.  Similar situations occur in
monitoring systems~\cite{pechenizkiy2010online}, fraud~\cite{vzliobaite2010adaptive}, and malware
detection~\cite{jordaney2017transcend,barbero2020transcending,yang2021cade}.

\paragraph{Explainable AI and drift explanation.} Explainable AI has recently
attempted to address the challenge of black-box model interpretation~\cite{ribeiro2016should, 
alvarez2018robustness, guo2018lemna}. Global
model-agnostic methods, such as Partial Dependence Plot
(PDP)~\cite{friedman2001greedy} and Accumulated Local Effects (ALE)
plot~\cite{apley2020visualizing, arzani2021interpretable}, provide visual clues on the effect of
different features on the prediction of black-box
models.  LEAF's LEAplot and LEAgram are inspired by PDP and ALE, but extend
these techniques by (1)~showing errors instead of effects; (2)~identifying the
correlated sets of features that contribute to drift, and (3)~helping to 
visualize how drift evolves over time in an operational setting. 
Recent research has developed techniques to explain concept drift in the
context of malware detection. Transcend~\cite{jordaney2017transcend} uses 
statistical comparison of samples to create decision boundary-based explanation.
CADE~\cite{yang2021cade} uses contrastive learning to develop a distance-based
explanation to find the feature sets that have the largest distance changes
towards the centroid in the latent space. Unfortunately, both explanation
approaches only apply to classification, and do not apply to prediction
in general, such as the regression-based prediction problems. To our knowledge, LEAF is the first framework to explain concept drift
for regression-based prediction problems for network management tasks.

\paragraph{Drift mitigation.} Adapting a model to migitate concept drift is
also a well-explored area~\cite{gama2014survey,lu2018learning}, but few mitigation 
approaches outperform frequent retraining~\cite{kantchelian2013approaches,thomas2011design,
mallick2022matchmaker,you2021learning}. Adaptation can
be based on retraining~\cite{xu2017dynamic} or model ensemble~\cite{krawczyk2017ensemble}.  Retraining-based approaches often
require complex data management and significant storage and memory. 
Paired Learners~\cite{bach2008paired} use one stable learner to learn
on old data and another reactive learner to learn from the latest data.
Accuracy Updated Ensemble
(AUE2)~\cite{brzezinski2011accuracy, brzezinski2013reacting} incrementally
updates sub-models on a small portion of data to adapt to drift. 
LEAF is inspired by several of these approaches, but focuses more heavily 
on identifying features that lead to concept drift, and performing 
mitigation at scale.

%% file: sections/conclusion.tex
\section{Conclusion}\label{sec:conclusion}

An important problem to operationalize machine learning
models to networking tasks in practice is concept drift. Although this phenomenon has 
been explored in other contexts, it has received limited attention in the 
networking domain. To address this critical problem, 
this paper characterizes drift patterns across multiple models and KPIs and has developed, 
presented, and evaluated LEAF, a framework to detect, explain, and mitigate drift 
for machine learning models applied to cellular demand forecasting. The LEAF framework employs 
explainable AI and informed mitigation. Our results based on more than 
four years of KPI data from this network show that LEAF consistently 
outperforms both periodic and triggered retraining while reducing the cost of 
retraining. 
We believe that the LEAF framework can be applied beyond cellular networks, to
other network management problems that use black-box models for regression-based 
prediction. Yet, these hypotheses are yet to be explored for future work. 
Another caveat is that the evaluation of LEAF to date 
has been conducted on fully labeled datasets; thus, a promising direction 
could be to improve LEAF to cope with semi-supervised or unsupervised models with 
partial or no labels.

%% file: sections/appendix.tex
\appendix

\begin{subappendices}



\section{Data and Drift Characteristics}\label{subsec:appedix_data}

\paragraph{\dataset.} Table~\ref{tab:target_KPIs_fixed} provides details of characteristics of target KPIs 
in \dataset. Still, it covers a variety of statistical patterns, which 
shows the diversity of this real-world dataset. Compared to \datasetlarge,
the dispersions (Std/Mean) of KPIs are less because no new sites are added. 

\begin{table}[h]
  \centering
  \small
  \begin{tabularx}{0.43\textwidth}{ccccccc}
    \toprule
    \multirow{3}{*}{\textbf{Property}} & \multicolumn{2}{c}{\textbf{Resource}}  
    & \multicolumn{2}{c}{\textbf{Network}} & \multicolumn{2}{c}{\textbf{User}}  \\ 
    & \multicolumn{2}{c}{\textbf{Utilization}}  
    & \multicolumn{2}{c}{\textbf{Performance}} & \multicolumn{2}{c}{\textbf{Experience}}\\
    \cmidrule(ll){2-3} \cmidrule(ll){4-5} \cmidrule(ll){6-7}
    & DVol & PU & DTP & REst   & CDR  & GDR  \\ \midrule
    Std/Mean & 0.73 & 1.34 & 0.57 & 0.77 & 1.35  & 2.12 \\ 
    \rowcolor{lightgray}
    Periodic & \Checkmark & \Checkmark & \Checkmark & \Checkmark & \Checkmark & \Checkmark \\ 
    Bursty &  & \Checkmark &  &  & \Checkmark & \Checkmark \\ 
    \rowcolor{lightgray}
    Data Lost &  & \Checkmark &  &  &  & \\ 
    Balanced & \Checkmark & \Checkmark & \Checkmark &  &  & \\ \bottomrule
  \end{tabularx}
  \caption{Characteristics of target KPIs in \dataset. DVol: Downlink volume; 
  PU: Peak active UEs; DTP: Downlink Throughput; REst: RRC establishment success; 
  CDR: S1-U call drop rate; GDR: RTP gap duration ratio.}
  \label{tab:target_KPIs_fixed}
\end{table}

\begin{figure}[h]
  \centering
  \subfloat[RRC Establishment.\label{fig:rrcest}]{%
  \includegraphics[width=0.245\textwidth]{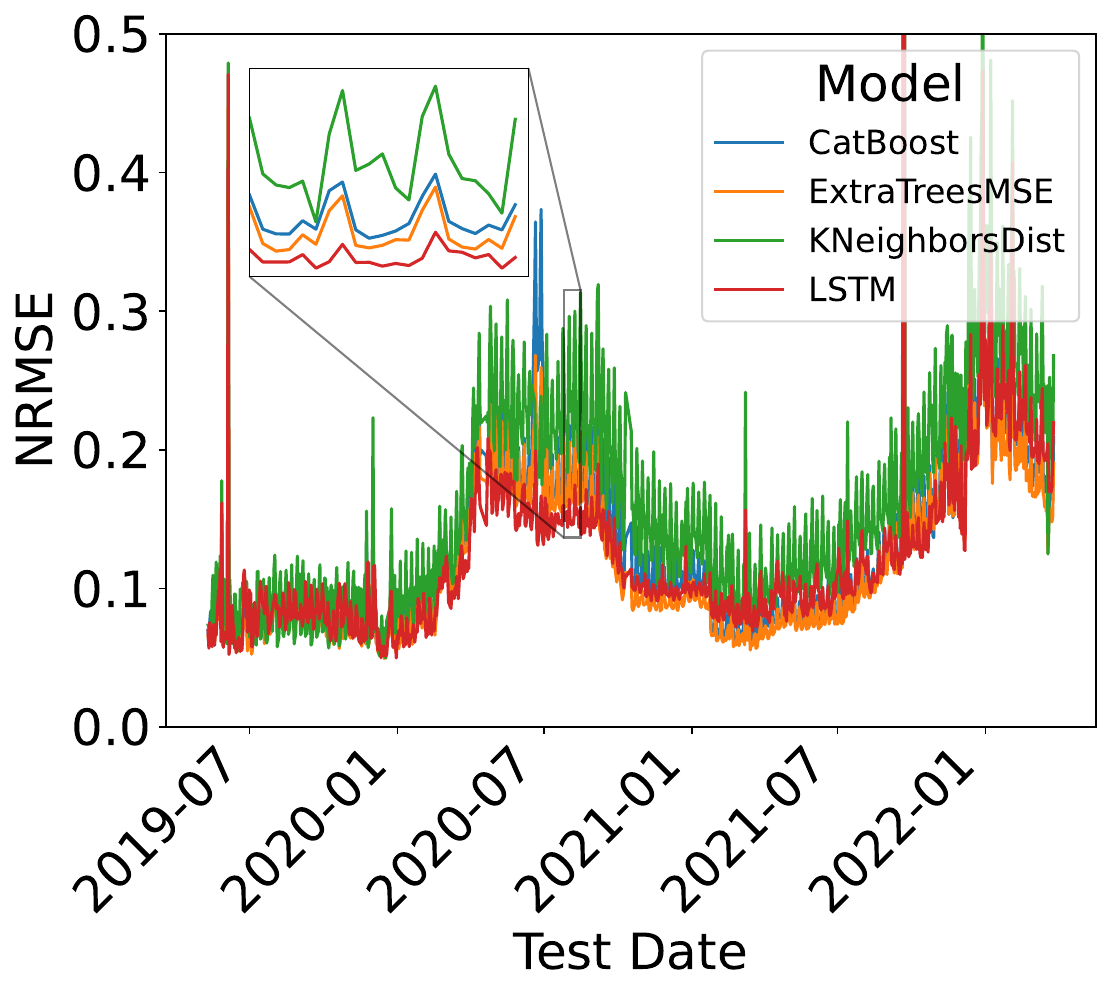}%
  }
  \subfloat[Call drop rate.\label{fig:calldrop}]{%
    \includegraphics[width=0.245\textwidth]{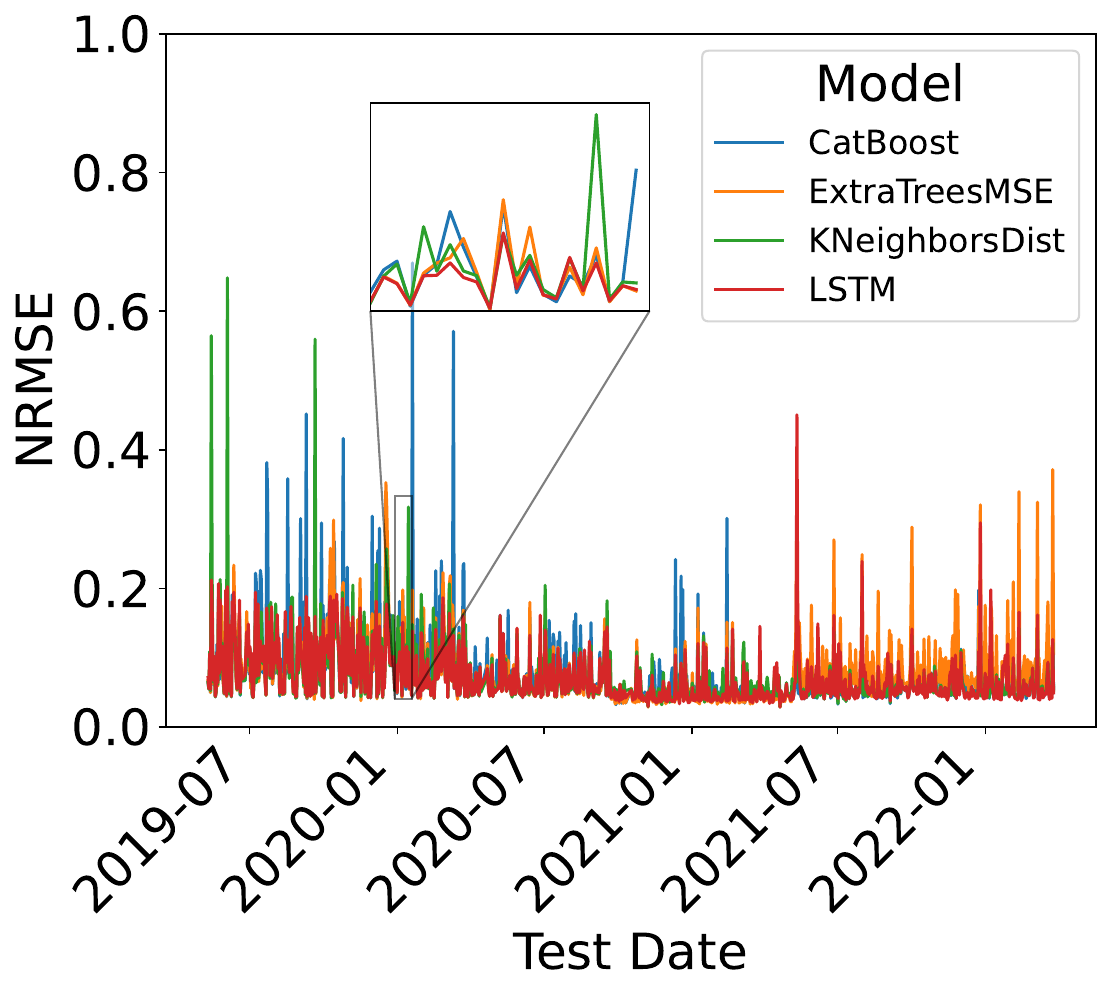}%
  }
  
  \caption{Drift of different models for KPIs of interest. Inset figures 
  exhibit a 3-week view (all starting from Sunday) of NRMSE for the box-selected 
  period.}
  \label{fig:kpi_concept_full}
\end{figure}

\paragraph{Drift across KPIs.} We add additional drift characterization plots of two more KPIs in 
Figure~\ref{fig:kpi_concept_full}, as it complements Figure~\ref{fig:kpi_concept}
in Section~\ref{sec:drift}. Models to predict RRC Establishment (REst) and 
Call Drop Rate (CDR) and other uncorrelated KPIs do not have the same drift 
patterns. Figure~\ref{fig:calldrop} exhibits higher values and larger 
fluctuations first, but the model stabilizes and shows improved NRMSE 
after April 2020. Moreover, short-lived, abrupt increases in error 
are more frequent than other KPIs, due to the burstiness of CDR.

\section{LEAF Details}\label{subsec:appedix_detection}

\paragraph{LEAF detector.} The drift detection module monitors the NRMSE time-series and notifies
operators when drift occurs. We feed NRMSE time-series into
Kolmogorov-Smirnov Windowing (KSWIN)~\cite{raab2020reactive,
togbe2021anomalies}. 
KSWIN is a recent concept drift detection method~\cite{raab2020reactive} based
on KS test, which is a non-parametric statistical test that makes no assumption
of underlying data distributions~\cite{togbe2021anomalies}. It monitors the
performance distributions as data arrives. We take the CatBoost NRMSE
time-series to forecast downlink volume as an example; the model is trained on a
total volume of 14 days of data before July 1, 2018 (illustrated in
Figure~\ref{fig:size}). Note that we do not have the ground truths of drifts in 
our dataset, thus we cross-check detection results with clear signals (\eg, 
missing data, network changes due to COVID-19). By applying KSWIN, we observe 
that instances of drift
are detected when the data exhibits major anomalies around June 2019, December
2019, and April 2021. The beginning and end of the COVID-19 quarantine period
are also effectively detected. We tested KSWIN across the NRMSE time-series of
the five KPIs of interest, using different model types and different training
set sizes and periods. KSWIN performs well across all tasks.

\begin{figure}[h]
  \centering
  \subfloat[Feature group 1.\label{fig:leaplot_g1}]{%
    \includegraphics[width=0.247\textwidth]{figures/leaplot_group1.pdf}%
  }
  \subfloat[Feature group 2.\label{fig:leaplot_g2}]{%
    \includegraphics[width=0.25\textwidth]{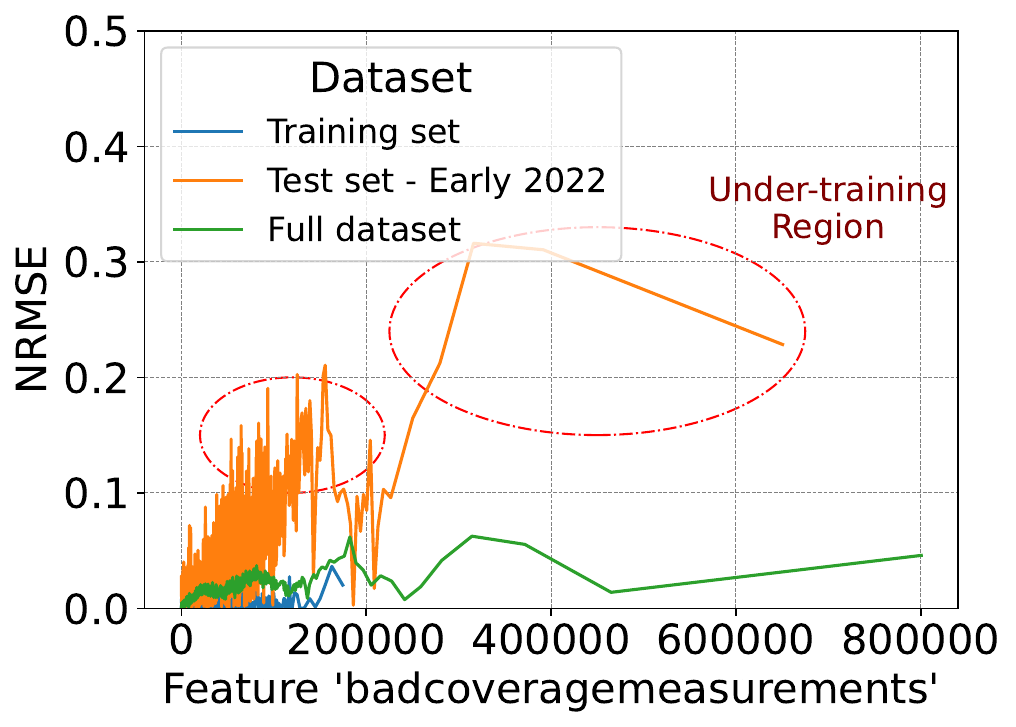}%
  }
  \caption{The LEAplots (1,000 bins) that decomposes CatBoost NRMSE time-series. The distribution of estimated local error is shown along with
  the values of the most representative features \code{pdcp\_dl\_datavol\_mb} 
  (downlink volume in Mbps, Group 1) and \code{badcoveragemeansurements} (number of 
  bad coverages, Group 2).}
  \label{fig:leaplot}
\end{figure}

\paragraph{LEAplot for different feature groups.} Figure~\ref{fig:leaplot}
shows the LEAplots of representative features from the top two feature groups of
our case study. In Figure~\ref{fig:leaplot_g1}, \code{pdcp\_dl\_datavol\_mb}
(the history of downlink volume) is the most representative feature from Group 1
for this model. Although the training set has a very low error when the feature
value changes, when its value is between 0.6e6 and 1.3e6, the errors in ``Early 2022'' test
set are more than 10x those of the training set and nearly 6x of those in the
full test set. For the value above 1.5e6, the training set does not cover the range 
so the test sets observe high errors on it. 
Figure~\ref{fig:leaplot_g2}, however, shows the LEAplots of representative
feature \code{badcoveragemeasurements} from Group 2. When the value is 
above 2e5, ``Early 2022'' test set has more errors than training set,
mainly because that the training set does not cover the range, and the model
is poorly fitted there.  

These two LEAplots exhibit very different error distributions. For example,
samples from the high error region of Figure~\ref{fig:leaplot_g1} are not 
always in that of Figure~\ref{fig:leaplot_g2}. It indicates that we can  
leverage this difference for iterative mitigation on different feature groups.

\section{End-to-end Mitigation Effectiveness}\label{subsec:appedix_e2e}
We implement the LEAF detection, explanation, and mitigation pipeline and
evaluate model performance before and after LEAF mitigation. We study how
LEAF's mitigation schemes improve model performance over the duration of the
constant eNodeB dataset. As the vast majority of errors occur in the tail end of
the distribution (see Figure~\ref{fig:directed_leagram}), we focus our study 
on the 95$^{th}$ percentile of errors. Further, we analyze the average error over
time to validate the impact that LEAF has on the model performance.

\begin{figure*}[!h]
  \begin{minipage}{0.3\textwidth}
    \begin{table}[H]
      \centering
      \begin{tabular}{crr}
        \toprule
        \multirow{2}{*}{\textbf{KPI}} & \multicolumn{2}{c}{\textbf{95$^{th}$ Error}} \\ \cmidrule(llll){2-3}
        & \textbf{Static} & \textbf{LEAF} \\ \midrule
          DVol & 0.29 & 0.19 \\ 
          PU & 0.86 & 0.27 \\ \midrule
          DTP & 0.17 & 0.13  \\
          REst & 0.33 & 0.18 \\\midrule
          CDR & 0.24 & 0.23\\
          GDR & 0.27 & 0.27\\\bottomrule
      \end{tabular}
      \caption{95$^{th}$ Normalized Error for CatBoost model on 6 
      KPIs (defined in Table~\ref{tab:target_KPIs}).}
      \label{tab:kpi_error}
  \end{table}
  \end{minipage}\hfill
  \begin{minipage}{0.7\textwidth}
    \begin{figure}[H]
      \centering
      \subfloat[Downlink Volume]{%
        \includegraphics[width=0.33\textwidth]{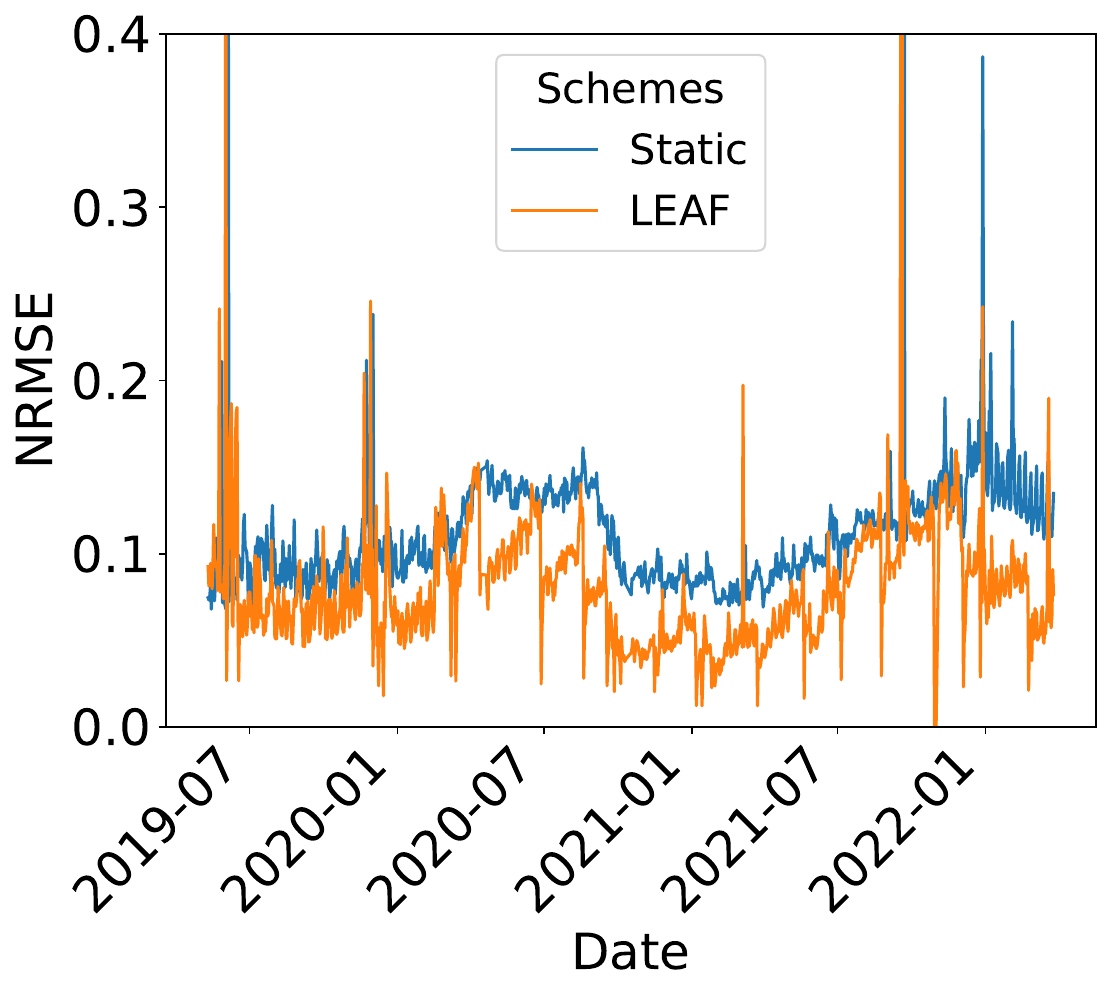}%
        \label{fig:DVol_mit}
      }
      \subfloat[Peak active User]{%
        \includegraphics[width=0.335\textwidth]{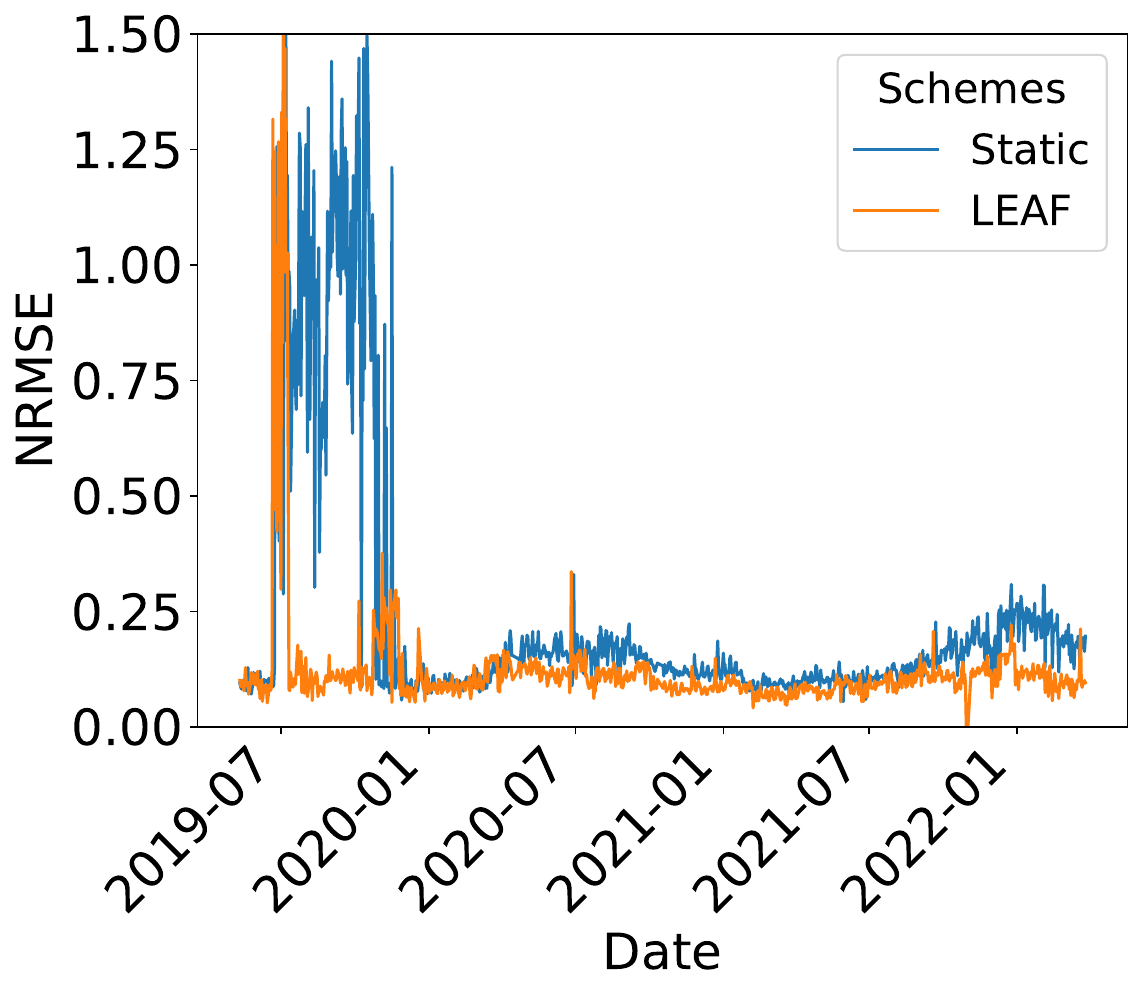}%
        \label{fig:PU_mit}
      }
      \subfloat[Downlink Throughput]{%
        \includegraphics[width=0.33\textwidth]{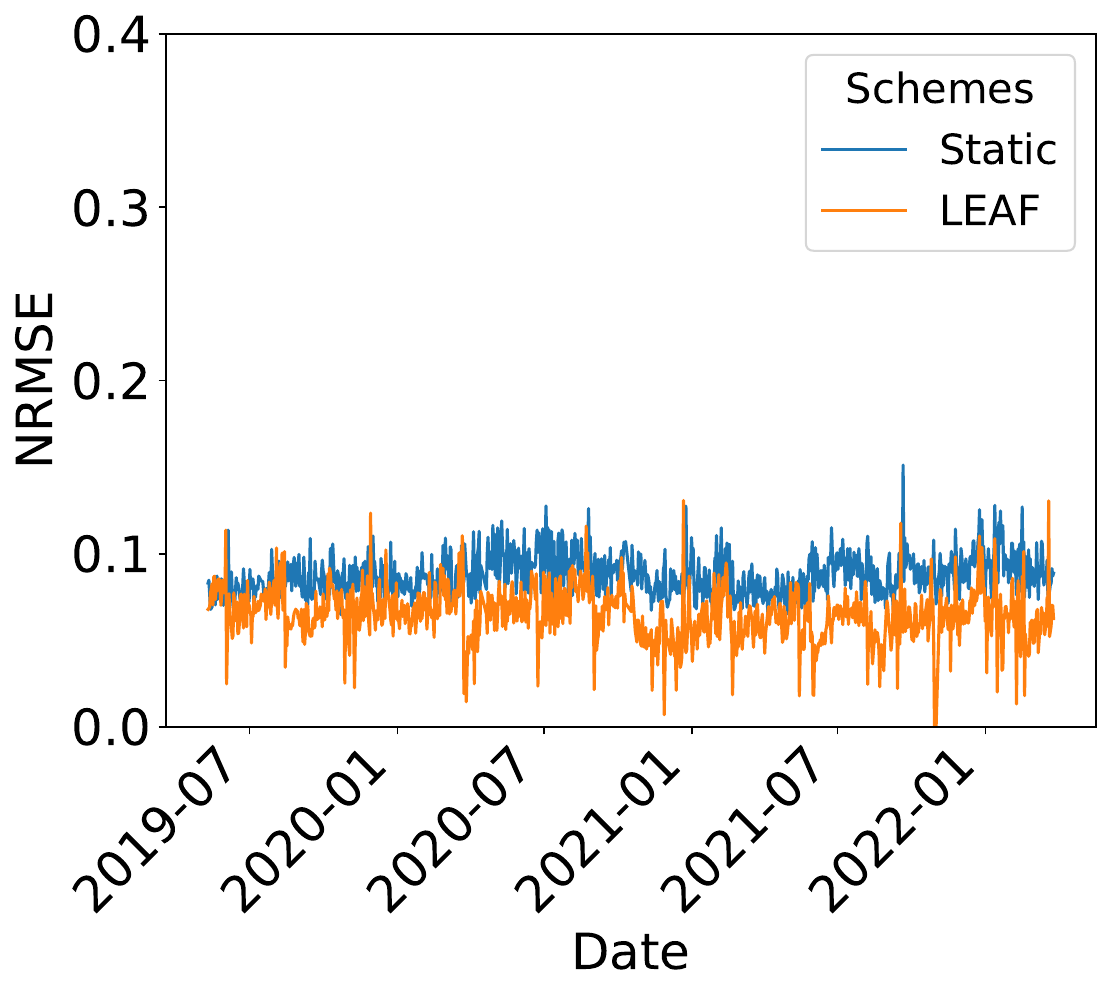}%
        \label{fig:DTP_mit}
      }\\
      \subfloat[RRC Establishmentatt]{%
        \includegraphics[width=0.33\textwidth]{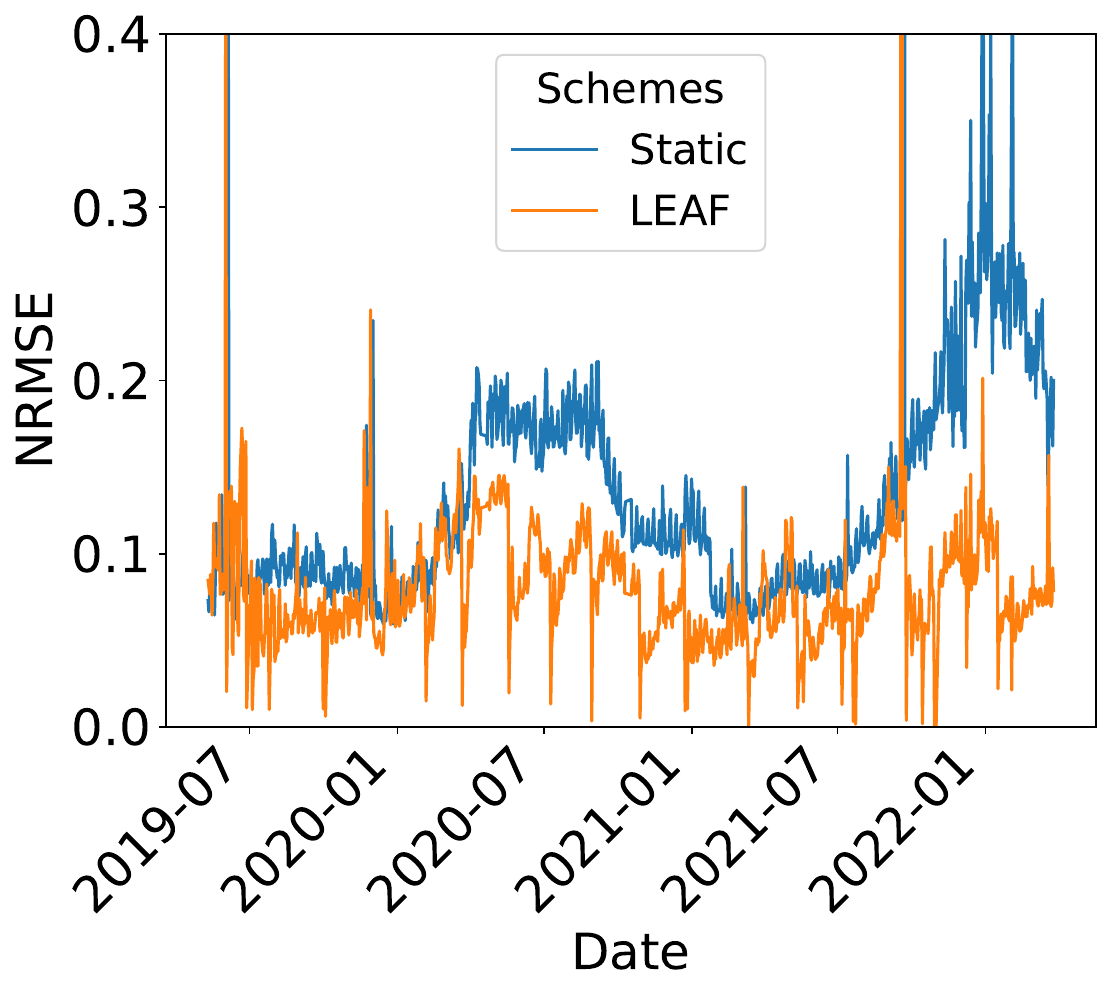}%
        \label{fig:REst_mit}
      }
      \subfloat[Call Drop Rate]{%
        \includegraphics[width=0.33\textwidth]{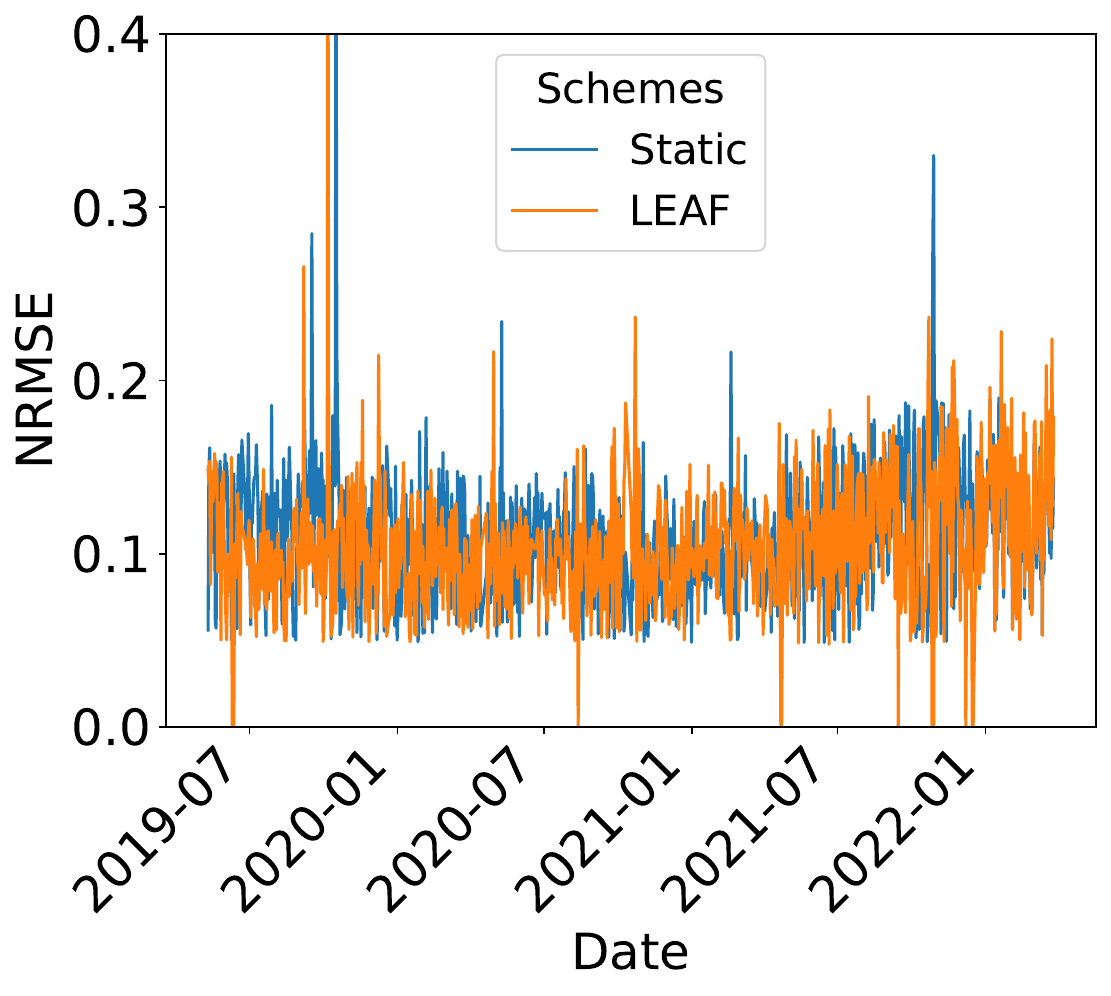}%
        \label{fig:CDR_mit}
      }
      \subfloat[Gap Duration Ratio]{%
        \includegraphics[width=0.33\textwidth]{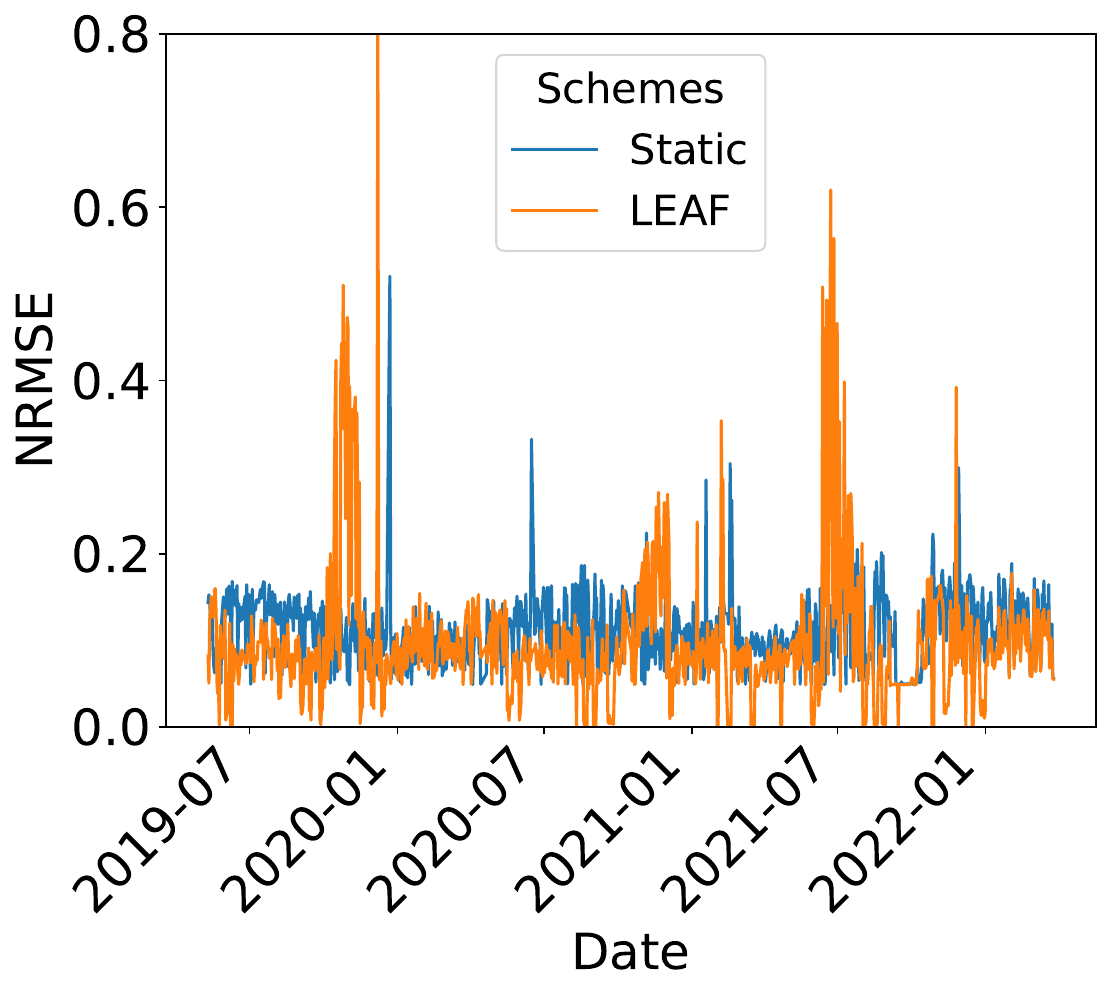}%
        \label{fig:GDR_mit}
      }
      \caption{NRMSE time-series before (static) and after (LEAF) mitigation, using LEAF.}
      \label{fig:e2e_mit}
      \vspace{10pt}
    \end{figure}
  \end{minipage}
\end{figure*}

\begin{figure*}[h]
  \centering
  \subfloat[Peak active UEs]{%
    \includegraphics[width=0.21\textwidth]{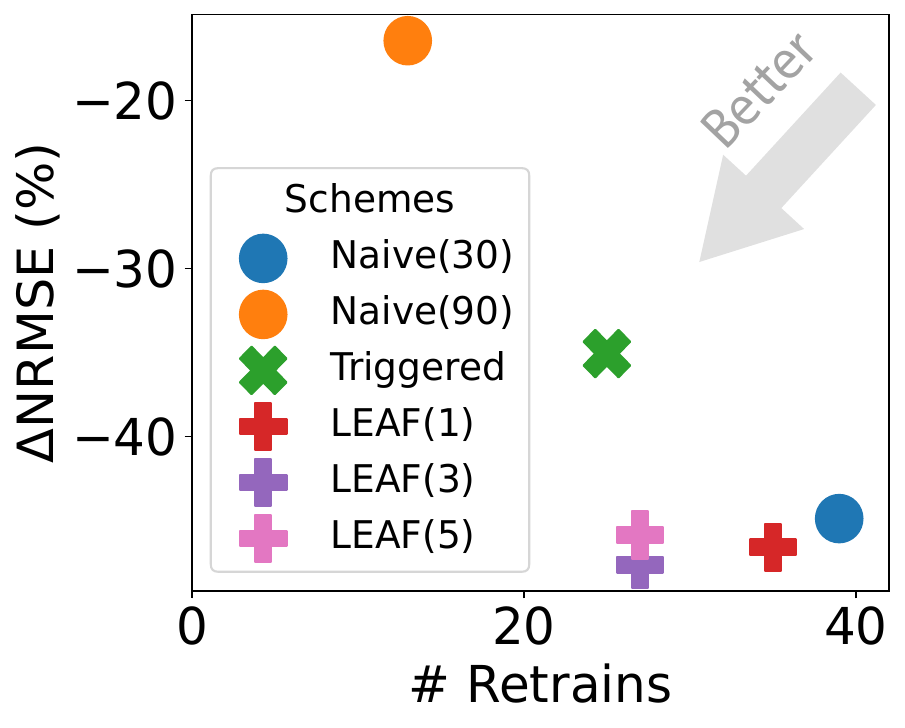}%
    \label{fig:PU_tradeoff}
  } 
  \subfloat[Downlink Throughput]{%
    \includegraphics[width=0.21\textwidth]{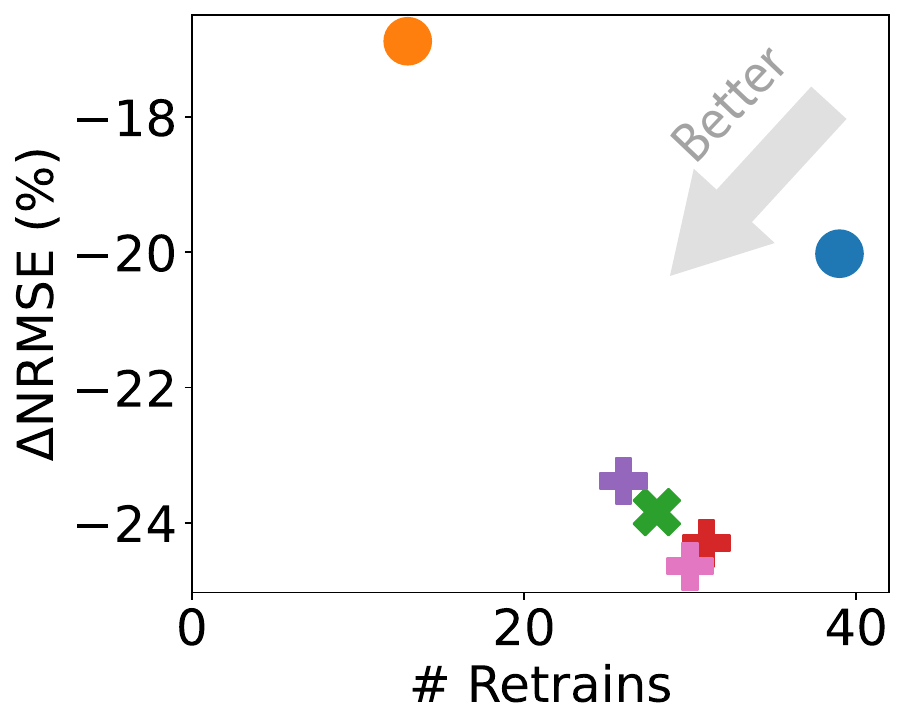}%
    \label{fig:DTP_tradeoff}
  } 
  \subfloat[RRC Establishmentatt]{%
    \includegraphics[width=0.21\textwidth]{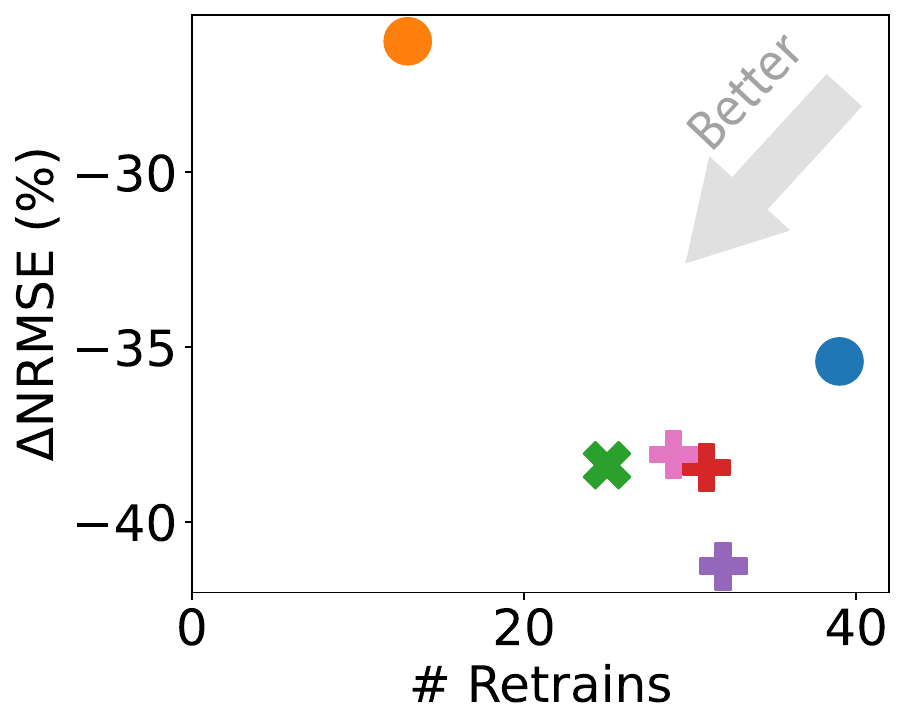}%
    \label{fig:REst_tradeoff}
  } 
  \subfloat[Gap Duration Ratio]{%
    \includegraphics[width=0.2\textwidth]{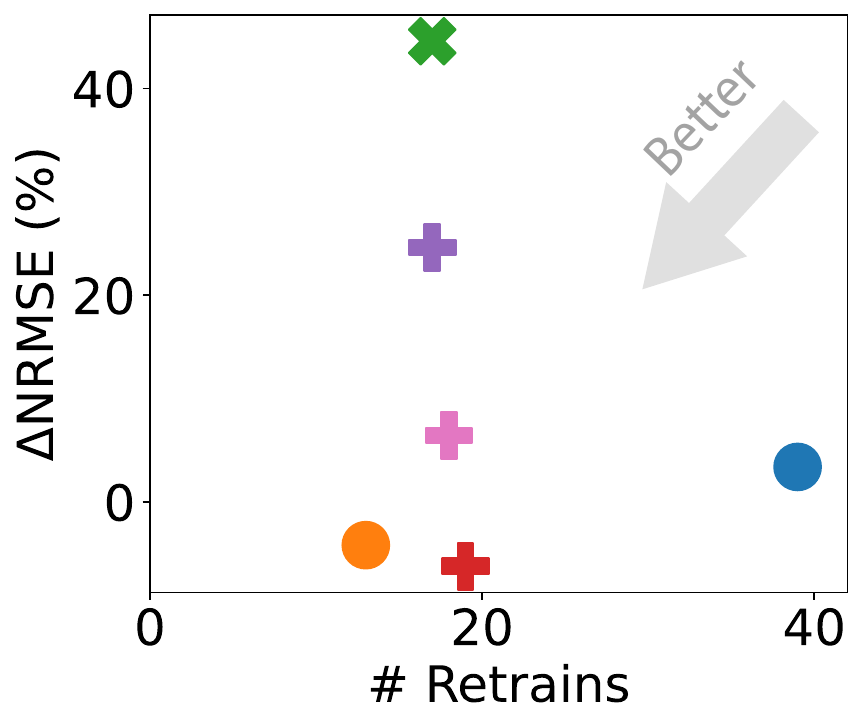}%
    \label{fig:GDR_tradeoff}
  }
  \caption{$\Delta \overline{NRMSE}$ vs. $\#Retrains$ under different 
  mitigation schemes using CatBoost. For na\"ive retraining, 
  number in parentheses denotes the retraining period in days. For LEAF,
  it represents the number of feature groups being used.
  Bottom left has the best mitigation effectiveness and lowest retraining
  cost.}
  \label{fig:e2e_tradeoff_full}
\end{figure*}

Table~\ref{tab:kpi_error} and Figure~\ref{fig:e2e_mit}, respectively, show the
95$^{th}$ percentile of the  
normalized error over the dataset and the daily NRMSE performance for the
CatBoost model when applied to 6 KPIs (defined in Table~\ref{tab:target_KPIs}).
We compare results obtained applying LEAF versus the static model (no
retraining). For each target KPI, we train a static model with the best
configuration obtained in Section~\ref{sec:drift}, \ie, 14 days of data before
July 1st, 2018. We apply LEAF on this baseline. Here, we show only results for
CatBoost as the overall behavior is consistent across different models. We
report detailed performance results across models and KPIs in the next
subsection.

We observe in Table~\ref{tab:kpi_error} that errors in the tail are largely
mitigated using LEAF on DVol, PU, DTP, and REst. For DVol, we observe that the
95$^{th}$ normalized error of CatBoost is effectively reduced from 0.29 for the
static model to 0.19 after LEAF is applied. PU is also significantly mitigated
by LEAF, reducing the 95$^{th}$ percentile of errors from 0.86 to 0.27.
However, CDR and GDR prove more difficult to mitigate possibly because that 
these two KPIs are highly dispersed (the coefficients of variance for them 
are 1.35 and 2.12 respectively, 2x to 4x higher compared to DVol, DTP, and 
REst). The errors obtained are only slight improvement over the baseline. 

In Figure~\ref{fig:e2e_mit} we study the NRMSE performance over time. We
observe that LEAF's mitigation consistently improves model performance compared
to the static model. The NRMSE obtained applying LEAF for DVol forecasting
(Figure~\ref{fig:DVol_mit}) never surpasses 0.125 (with the exception of spikes
due to data errors) and goes as low as 0.02. This ensures a bounded error
distribution and decreases the average error by 32.67\%. We also observe that
mitigation promptly occurs once drift is detected. In Figure~\ref{fig:DVol_mit}
and Figure~\ref{fig:REst_mit}, this is particularly evident for sudden changes
in the NRMSE distribution (\eg, around January and April 2020 because of
COVID-19 lockdowns) and gradual changes (\eg, from the second half of 2021 to
the beginning of 2022). Similarly, for PU forecast we observe in
Figure~\ref{fig:PU_mit} that large unexpected errors are promptly mitigated by
LEAF. High errors exceeding an NRMSE of 1 caused by missing data are observed
during only a week (around June 2019), in contrast to the static model that
continues to exhibit high errors for more than half a year. Finally, While LEAF
still slightly improves model performance for highly dispersed KPIs like CDR
(Figure~\ref{fig:CDR_mit}) and GDR (Figure~\ref{fig:GDR_mit}), they prove more
difficult to mitigate. For example, due to the bursty nature of GDR, LEAF 
exhibits two bursty error spikes in Figure~\ref{fig:GDR_mit}, around December 
2020 and August 2021, while all other KPIs do not show this pattern. 

\section{Full Evaluation}\label{subsec:appedix_full_eval}

We provide supplementary results as it shows the full evaluation across 
all KPIs, models, and mitigation schemes. Figure~\ref{fig:e2e_tradeoff_full}
complements Figure~\ref{fig:e2e_tradeoff} for the remaining KPIs. It 
illustrates $\Delta \overline{NRMSE}$ vs. $\#Retrains$ under different 
mitigation schemes using CatBoost. Table~\ref{tab:delta_nrmse_full} and 
~\ref{tab:delta_nrmse_evolv_full} show the full evaluation on effectiveness 
of all mitigation schemes measured in $\Delta \overline{NRMSE}$ and $\#Retrains$ 
using \dataset and \datasetlarge respectively.

\begin{table*}[h]
  \centering
\newcolumntype{s}{>{\hsize=.4\hsize}r}
\begin{tabularx}{0.73\textwidth}{ccrlrlrlrl}
  \toprule
  \multirow{2}{*}{\textbf{Model}} & \multirow{2}{*}{\textbf{KPIs}} & \multicolumn{8}{c}{\textbf{$\Delta \overline{NRMSE}$ ($\#$Retrains)  of Mitigation Schemes}}  \\ 
  \cmidrule(llll){3-10}
   &  & \multicolumn{2}{c}{Na\"ive$_{30}$} & \multicolumn{2}{c}{Na\"ive$_{90}$} & \multicolumn{2}{c}{Triggered} & \multicolumn{2}{c}{LEAF} \\ \midrule
   \multirow{6}{*}{\textbf{CatBoost}} & DVol & $-29.62\%$ & (39) & $-19.83\%$ & (13) & $-31.80\%$ & (27) & \cellcolor[HTML]{b7b7b7}$-32.67\%$ & (28) \\
   & PU & $-44.88\%$ & (39) & $-16.44\%$ & (13) & $-35.06\%$ & (25) & \cellcolor[HTML]{b7b7b7}$-46.59\%$ & (35) \\
   & DTP & $-20.02\%$ & (39) & $-16.88\%$ & (13) &  $-23.84\%$ & (28) & \cellcolor[HTML]{b7b7b7}$-24.30\%$ & (31) \\
   & REst & $-35.41\%$ & (39) & $-26.25\%$ & (13) & $-38.38\%$ & (25) & \cellcolor[HTML]{b7b7b7}$-38.44\%$ & (31) \\
   & CDR & $2.35\%$ & (39) & \cellcolor[HTML]{b7b7b7}$-5.39\%$ & (13) & $-4.21\%$ & (17) & $-3.63\%$ & (9) \\
   & GDR & $3.37\%$ & (39) & $-4.20\%$ & (13) & $44.56\%$ & (17) & \cellcolor[HTML]{b7b7b7}$-6.24\%$ & (19) \\ \midrule
   \multirow{6}{*}{\textbf{ExtraTrees}} & DVol & $-24.77\%$ & (39) & $-15.10\%$ & (13) & $-28.17\%$ & (32) & \cellcolor[HTML]{b7b7b7}$-30.64\%$ & (32) \\
   & PU & $-44.26\%$ & (39) & $-16.30\%$ & (13) & \cellcolor[HTML]{b7b7b7}$-50.76\%$ & (26) & $-45.83\%$ & (27) \\
   & DTP & $-18.13\%$ & (39) & $-13.80\%$ & (13) & $-21.63\%$ & (32) & \cellcolor[HTML]{b7b7b7}$-22.59\%$ & (23) \\
   & REst & $-31.95\%$ & (39) & $-22.28\%$ & (13) & $-34.29\%$ & (22) & \cellcolor[HTML]{b7b7b7}$-36.13\%$ & (29) \\
   & CDR & $2.10\%$ & (39) & \cellcolor[HTML]{b7b7b7}$-2.52\%$ & (13) & $8.08\%$ & (20) & $-0.20\%$ & (11) \\
   & GDR & $-0.58\%$ & (39) & $9.64\%$ & (13) & $33.67\%$ & (17) & \cellcolor[HTML]{b7b7b7}$-14.26\%$ & (19) \\ \midrule
   \multirow{6}{*}{\textbf{LSTM}} & DVol & $0.54\%$ & (39) & \cellcolor[HTML]{b7b7b7}$-0.97\%$ & (13) & $14.12\%$ & (21) & $2.67\%$ & (19) \\
   & PU & $37.11\%$ & (39) & $12.65\%$ & (13) & $3.76\%$ & (25) & \cellcolor[HTML]{b7b7b7}$-20.48\%$ & (18) \\
   & DTP & $17.08\%$ & (39) & $11.94\%$ & (13) & $-0.82\%$ & (14) & \cellcolor[HTML]{b7b7b7}$-37.13\%$ & (20) \\
   & REst & $6.78\%$ & (39) & \cellcolor[HTML]{b7b7b7}$3.57\%$ & (13) & $5.78\%$ & (27) & $4.21\%$ & (26) \\
   & CDR & $-33.22\%$ & (39) & $-56.84\%$ & (13) & $-21.39\%$ & (10) & \cellcolor[HTML]{b7b7b7}$-71.52\%$ & (11) \\
   & GDR & $0.41\%$ & (39) & $-0.53\%$ & (13) & $-8.58\%$ & (14) & \cellcolor[HTML]{b7b7b7}$-16.29\%$ & (13) \\ \midrule
   \multirow{6}{*}{\textbf{KNeighbors}} & DVol & \cellcolor[HTML]{b7b7b7}$-8.26\%$ & (39) & $-2.99\%$ & (13) & $-4.11\%$ & (16) & $-4.47\%$ & (24) \\
   & PU & $-34.09\%$ & (39) & $-8.08\%$ & (13) & \cellcolor[HTML]{b7b7b7}$-37.99\%$ & (16) & $-18.11\%$ & (20) \\
   & DTP & \cellcolor[HTML]{b7b7b7}$-4.73\%$ & (39) & $-2.52\%$ & (13) & $-4.03\%$ & (18) & $-1.53\%$ & (22) \\
   & REst & \cellcolor[HTML]{b7b7b7}$-26.69\%$ & (39) &  $-21.74\%$ & (13) & $-25.86\%$ & (25) & $-22.10\%$ & (16) \\
   & CDR & $9.44\%$ & (39) & \cellcolor[HTML]{b7b7b7}$2.50\%$ & (13) & $7.35\%$ & (11) & $4.69\%$ & (12) \\
   & GDR & $-8.13\%$ & (39) & $-21.16\%$ & (13) & \cellcolor[HTML]{b7b7b7}$-23.40\%$ & (19) & $-6.12\%$ & (13) \\ \bottomrule
  \end{tabularx}
  \caption{Effectiveness of mitigation schemes measured in $\Delta \overline{NRMSE}$ and $\#Retrains$ 
  (both are the lower the better) using \dataset. We include representative 
  models from different model families over a variety of KPIs. }
  \label{tab:delta_nrmse_full}
\end{table*}

\begin{table*}[h]
  \centering
\newcolumntype{s}{>{\hsize=.4\hsize}r}
\begin{tabularx}{0.86\textwidth}{ccrlrlrlrlrl}
  \toprule
  \multirow{2}{*}{\textbf{Model}} & \multirow{2}{*}{\textbf{KPIs}} & \multicolumn{10}{c}{\textbf{$\Delta \overline{NRMSE}$ ($\#$Retrains) of Mitigation Schemes}}  \\ 
  \cmidrule(llll){3-12}
   &  & \multicolumn{2}{c}{Na\"ive$_{30}$} & \multicolumn{2}{c}{Na\"ive$_{90}$} & \multicolumn{2}{c}{Triggered} & \multicolumn{2}{c}{LEAF} & \multicolumn{2}{c}{LEAF*}\\ \midrule  
   & DVol & $-29.62\%$ & (39) & $-19.83\%$ & (13) & $-31.80\%$ & (27) & $-32.67\%$ & (28) & \cellcolor[HTML]{b7b7b7}$-35.12\%$ & (34) \\
   & PU & $-44.88\%$ & (39) & $-16.44\%$ & (13) & $-35.06\%$ & (25) & $-46.59\%$ & (35) & \cellcolor[HTML]{b7b7b7}$-47.62\%$ & (27) \\
   \textbf{Fixed} & DTP & $-20.02\%$ & (39) & $-16.88\%$ & (13) &  $-23.84\%$ & (28) & $-24.30\%$ & (31) & \cellcolor[HTML]{b7b7b7}$-24.64\%$ & (30) \\
   \textbf{Dataset} & REst & $-35.41\%$ & (39) & $-26.25\%$ & (13) & $-38.38\%$ & (25) & $-38.44\%$ & (31) & \cellcolor[HTML]{b7b7b7}$-41.27\%$ & (32)\\
   & CDR & $2.35\%$ & (39) & $-5.39\%$ & (13) & $-4.21\%$ & (17) & $-3.63\%$ & (9) & \cellcolor[HTML]{b7b7b7}$-6.22\%$ & (12)\\
   & GDR & $3.37\%$ & (39) & $-4.20\%$ & (13) & $44.56\%$ & (17) & \cellcolor[HTML]{b7b7b7}$-6.24\%$ & (19) & \cellcolor[HTML]{b7b7b7}$-6.24\%$ & (19) \\ \midrule
       
   & DVol & $-29.00\%$ & (39) & $-19.53\%$ & (13) & $-30.76\%$ & (24) & $-32.09\%$ & (37) & \cellcolor[HTML]{b7b7b7}$-32.80\%$ & (30)\\
   & PU & $-44.56\%$ & (39) & $-17.38\%$ & (13) & $-50.89\%$ & (24) & $-45.75\%$ & (24) & \cellcolor[HTML]{b7b7b7}$-51.72\%$ & (26)\\
   \textbf{Evolving} & DTP & $-19.51\%$ & (39) & $-17.59\%$ & (13) & $-22.19\%$ & (30) & \cellcolor[HTML]{b7b7b7}$-22.58\%$ & (27) & \cellcolor[HTML]{b7b7b7}$-22.58\%$ & (27)\\
   \textbf{Dataset} & REst & $-37.51\%$ & (39) & $-28.33\%$ & (13) & $-44.01\%$ & (25) & $-43.66\%$ & (26) & \cellcolor[HTML]{b7b7b7}$-48.01\%$ & (33) \\
   & CDR & $-3.79\%$ & (39) & $-1.24\%$ & (13) & $-6.79\%$ & (7) & $-1.33\%$ & (9) & \cellcolor[HTML]{b7b7b7}$-7.15\%$ & (8) \\
   & GDR & $-8.46\%$ & (39) & $-2.65\%$ & (13) & \cellcolor[HTML]{b7b7b7}$-13.21\%$ & (15) & $-2.06\%$ & (13) & $-11.99\%$ & (17)\\ \bottomrule
  \end{tabularx}
  \caption{Effectiveness of different mitigation schemes measured in $\Delta \overline{NRMSE}$ and $\#Retrains$ 
  (both are the lower the better) using both datasets. We show the best scheme of 
  multi-group LEAF and denote it as LEAF*}
  \label{tab:delta_nrmse_evolv_full}
\end{table*}

\end{subappendices}